\documentclass[galaxies,article,accept,pdftex,moreauthors]{Definitions/mdpi} 
\usepackage{graphicx}
\usepackage{subcaption}
\firstpage{1} 
\pubvolume{1}
\issuenum{1}
\articlenumber{0}
\pubyear{2024}
\copyrightyear{2024}
\datereceived{11 December 2023} 
\daterevised{1 March 2024} % Comment out if no revised date
\dateaccepted{7 March 2024} 
\datepublished{ } 
\hreflink{https://doi.org/} % If needed use \linebreak
\Title{Central Engine and Spectral Energy Distribution Properties of High Redshift Gamma Ray Blazars}
\TitleCitation{Central Engine and Spectral Energy Distribution Properties of High Redshift Gamma Ray Blazars}
\Author{Anilkumar Tolamatti %MDPI: Please carefully check the accuracy of names and affiliations.
 $^{1,2,}$*\orcidA{}, Krishna Kumar Singh $^{1,2,}$\orcidB{} and Kuldeep Kumar Yadav $^{1,2}$ }

\AuthorNames{Anilkumar Tolamatti, Krishna Kumar Singh and Kuldeep Kumar Yadav}

\AuthorCitation{Tolamatti, A.; Singh, K.K.; Yadav, K.K.}
\address{%
$^{1}$ \quad Astrophysical Sciences Division, Bhabha Atomic Research Centre, Mumbai 400085, India; kksastro@barc.gov.in\\
$^{2}$ \quad Homi Bhabha National Institute, Anushakti Nagar, Mumbai 400094, India}

\corres{Correspondence:anilkumart@barc.gov.in~(A.T.)}

\abstract{We report on the properties of central engines in the $\gamma$-ray blazars located at high redshifts 
beyond z~>~0.4, where the extra-galactic background light (EBL) starts affecting their $\gamma$-ray spectra.
The physical engine that provides power to the blazars of very high bolometric luminosity is 
assumed to be a highly collimated jet of matter moving relativistically away from the supermassive 
black hole (SMBH), located in the central region of the host galaxy, in a direction aligned toward the Earth. 
Due to their peculiar geometry and special physical conditions, blazars at redshifts beyond z~>~0.4 are 
bright enough to be detected in the $\gamma$-ray energy band. In this work, we investigate the physical properties of 
high-$z$ $\gamma$-ray blazars detected by the Large Area Telescope (LAT) on board the \emph{Fermi} satellite. 
We also study the properties of their emission regions and the central engines and discuss cosmological and astrophysical implications.}


\begin{document}

%%%%%%%%%%%%%%%%%%%%%%%%%%%%%%%%%%%%%%%%%% Introduction %%%%%%%%%%%%%%%%%%%%%%%%%%%%%%%%%%%%%%%

\section{Introduction}
Blazars are bright beacons in the Universe. They represent a subclass of the radio-loud active galactic nuclei (AGNs) which 
are characterized as elliptical galaxies hosting supermassive black holes (SMBHs;$M_{BH}:10^6-10^{10} M_\odot$) at their 
centers and oppositely oriented powerful jets of plasma ejected from the poles of SMBH~\cite{Urry1995,Padovani2017}. 
The accretion of gaseous matter onto the spinning SMBH is considered to be the dominant physical process for powering these sources 
and launching the relativistic plasma jets. However, the exact mechanism of jet formation is not completely understood and 
three alternative theories have been proposed. In the so-called Blandford--Znajek (BZ) mechanism, the jet extracts energy from the 
rotation of SMBH, and the power of the jet depends on the black hole spin and square of magnetic flux threading its event 
horizon~\cite{Blandford1977}. In the second model, also known as the Blandford--Payne (BP) mechanism, the jet extracts energy only 
from the rotation of the accretion disk, and the role of SMBH is not important~\cite{Blandford1982}. In the case of these two theories, the formation 
of jets is sustained by the accretion of matter onto the SMBHs, and a relation between jet and accretion luminosity is expected and 
has been confirmed in different studies~\cite{Maraschi2003,Ghisellini2011,Chen2015,Zhang2022}. The hybrid model, a combination 
of the above two mechanisms, speculates about the observed differences in different types of AGNs with jets~\cite{Meier1999}. The jets in 
AGNs are observed to be ultra-luminous sources of non-thermal radiation spanning over the entire electromagnetic spectrum ranging 
from radio waves to very high energy (VHE; E $>$ 20 GeV) $\gamma$-rays. They are also thought to be the apparent manifestations of 
AGN activity during their early evolutionary stages~\cite{Komissarov2021}. An extreme class of such AGNs, in which one of the jets 
points close to the line of sight of the observer, is referred to as blazars. Due to the peculiar orientation of the blazar jets, 
their emission undergoes relativistic Doppler boosting and beaming~\cite{Urry1995,Padovani2017}. As a consequence, the intrinsic emission 
from blazars is strongly enhanced and the characteristic timescales are shortened for an observer on Earth. As a result, blazars 
are observed to exhibit distinct observational features such as rapid and large amplitude flux variability on most timescales across 
the electromagnetic spectrum, spectral changes (harder-when-brighter), strong and variable optical polarization, variable high 
energy (HE; E $>$ 30 MeV) $\gamma$-ray emissions, and apparent superluminal motion~\cite{Wagner1995,Giommi2021,Lister2019,Weaver2020,
Shukla2020,Singh2020a,Singh2020b,Tolamatti2022}. The intensity amplification effects make even faint blazars detectable in the 
multi-wavelength observations. Thus, blazars are the ideal laboratories to probe the physics properties of AGNs and study the growth 
and evolution of SMBHs using the measurements of their broadband emission~\cite{Singh2022}. 
\par
\textls[-15]{The broadband spectral energy distribution (SED) of blazars typically exhibits two characteristic broad humps. 
The first hump,  peaking at low energies in radio to UV/soft X-ray,  is produced by the synchrotron emission from 
the population of relativistic electrons in the presence of jet magnetic field. The origin of the second hump, which peaks 
in the $\gamma$-ray band is not well understood. It is often attributed to inverse Compton (IC) up-scattering of low energy 
seed photons by the relativistic jet electrons in the emission region. The seed photons for the IC process may be the synchrotron 
photons produced within the jet (synchrotron self Compton or SSC~\cite{Marscher1985}) or thermal photons (UV/IR) originating 
from various components such as accretion disk, broad line region (BLR) and dusty torus (DT) external to the jet. These processes 
are referred to as external Compton (EC) for HE and VHE emissions in blazars~\cite{Sikora1994,Ghisellini2009}. Alternatively, hadronic models 
involving the ultra-relativistic protons and higher jet magnetic field are invoked to explain the observed VHE emission through proton synchrotron, 
photo-hadronic interactions, and subsequent meson decay~\cite{Aharonian2000,Bottcher2013,Petropoulou2015}.}
\par 
\textls[-15]{Blazars are historically classified into two categories: Flat Spectrum Radio Quasars (FSRQs) and BL Lacertae Objects (BL Lacs), on 
the basis of strength, visibility, and rest-frame equivalent width (EW) of the emission/absorption lines in their optical continuum~\cite{Stickel1991}. 
FSRQs are characterized by strong and broad emission lines with EW $> 5$ \AA ~ while BL Lacs have 
typically weak or no emission lines. Another classification of blazars is also proposed on the basis of the known features of 
double-hump structures in their broadband SEDs. The position of the low energy hump or synchrotron peak distinguishes the blazars 
as Low Synchrotron Peak (LSP), Intermediate Synchrotron Peak (ISP), and  High Synchrotron Peak (HSP) objects~\cite{Abdo2010,Padovani2012,Giommi2012}. 
For LSP blazars, the synchrotron peak frequency of $ \leq 10^{14}$ Hz, corresponds to the radio to infrared regime. ISP blazars have synchrotron 
peak frequency in the range $10^{14}$--$10^{15}$ Hz, which lies in the optical/UV regime. Finally, HSP blazars exhibit the synchrotron peak 
frequency above $10^{15}$ Hz, corresponding to the X-ray regime. Further, the high-energy hump or Compton peak of BL Lacs is equally luminous as 
the low-energy synchrotron peak and is attributed to the SSC process. Whereas in the case of FSRQs, the Compton peak is more luminous than the 
synchrotron peak and is generally ascribed by the EC process. Also, the accretion disc in FSRQs is more radiatively efficient than that 
in BL Lacs~\cite{Ghisellini2011,Padovani2017}. This provides a more physical distinction between the blazar subclasses. The population 
study of a large sample of blazar SEDs indicates a so-called \emph{blazar sequence}, which characterizes an empirical inverse relationship 
between the bolometric $\gamma$-ray luminosity and the synchrotron peak frequency~\cite{Fossati1998,Prandini2022}. This implies that the synchrotron 
peak frequency is anti-correlated with the ratio of Compton to synchrotron peak (Compton dominance), i.e., as the bolometric luminosity increases 
synchrotron peak frequency decreases. According to this, FSRQs exhibit the lowest synchrotron peak frequency and highest bolometric luminosity whereas 
HSP blazars have the lowest bolometric luminosity. It means FSRQs and BL Lacs can be quantified as high and low-power blazars, respectively. 
Also, this empirical observation feature suggests that accretion and blazar jet powers are related, and BL Lacs with heavier black holes and FSRQs 
with lighter black holes have the same $\gamma$-ray luminosity~\cite{Ghisellini2016,Ghisellini2017,Singh2019a}. The origin of the blazar sequence has not 
been fully understood yet and is generally attributed to the differences in the electron cooling efficiency or 
Doppler boosting~\cite{Ghisellini1998,Fan2017}.  }
\par
In recent years, a class of high-$z$ blazars has acquired particular importance in the field of blazar research \cite{Paliya2020,Sahakyan2020}. 
The GeV--TeV photons emitted from these sources suffer from partial attenuation during their propagation due to the interaction with low energy 
extragalactic background light (EBL) via $\gamma-\gamma$ pair production~\cite{Stecker2006,Singh2020c}. In this study, we mainly focus on the 
investigation of the physical properties of high-$z$ blazars using publicly available data from broadband observations and SED modeling in 
the literature. The paper is organized as follows: A brief overview of the high redshift $\gamma$-ray blazars is given in Section~\ref{high-z}. 
In Section~\ref{data}, we describe the data sample used in this work and discuss the results in Section~\ref{results}. 
A summary of this study and future prospects of high-$z$ blazars are presented in Section~\ref{summary}.

%---------------------------------Section-2-High Redshift-Blazars--------------------------------------------------------------------
\section{High-$z$ Blazars}\label{high-z}
We use a purely arbitrary definition of $z > 0.4$, in the present work, for characterizing the high redshift $\gamma$-ray blazars in the sky. 
However, this definition can be supported by the phenomenological evidence that beyond  $z > 0.4$ the imprints of EBL absorption start becoming 
dominant in the HE $\gamma$-ray spectra measured by the Large Area Telescope on the Fermi spacecraft (\emph{Fermi}-LAT) in the energy range 
above few tens of GeV~\cite{Tolamatti2022a}. At these redshifts, the effects of cosmological evolution on structure formation and HE $\gamma$-ray 
propagation also becomes important. Thus, high-$z$ $\gamma$-ray blazars detected by the \emph{Fermi}-LAT are ideal tools to probe the early epochs 
in the evolution of the Universe as their bolometric luminosity is dominated by the HE emission~\cite{Ghisellini2013,Ackermann2015,Paliya2020,Sahakyan2020,
Inayoshi2020,Belladitta2022,Ajello2014}. The host galaxies of these sources are generally observed to host extremely massive black holes 
($M_{BH} > 10^9 M_\odot$) at their center. Therefore, high-$z$ blazars can constrain the mass function at the high end of SMBHs and provide a theoretical 
understanding of the growth and evolution of black holes over cosmic time~\cite{Johnson2016}.  Thus, on account of being a bright HE $\gamma$-ray source, 
blazars at high redshift are ideal objects for exploring the energetics and emission mechanisms under extreme conditions, setting strong constraints 
on the EBL intensity, and shedding light on the formation and growth of SMBHs in the early Universe and also their connection with the powerful 
relativistic jets. The detection and understanding of the physical properties of a few high-$z$ blazars in a given redshift bin enable us to infer the 
total density and behavior of radio-loud AGNs in that redshift bin because for each blazar with the Bulk Lorentz factor ($\Gamma~\approx$10--20) of 
the plasma within the jet, there must exist $2\Gamma^2$ sources sharing the same properties but appearing different due to their jet pointing 
elsewhere~\cite{Ghisellini2013}. This complements the independent estimation based on AGNs with misaligned jets. Therefore, it is important to identify 
and study the high redshift blazars to understand the cosmic evolution of SMBHs. Recent radio observations have reported the discovery of the most 
distant blazar $PSO~J0309+27$ located at a redshift of $z~\approx$ 6.1 when the age of the Universe was less than $\approx$ 1 billion years~\cite{Belladitta2020}. 
Its central engine is found to be a black hole of mass 1.45$_{-0.89}^{+1.89} \times 10^{9}$ M$_{\odot}$  with a bolometric luminosity 
of $\approx ~8 \times10^{46}$ erg~s$^{-1}$. This corresponds to a seed black hole mass of 10$^6 M_\odot$ at $z~=~30$ (when the first stars and galaxies are 
thought to have formed $\approx$ 15 million years after the Big Bang)~\cite{Belladitta2022}. Such a rapid growth of SMBHs from the early stars has  
serious implications on various evolution models for the SMBHs in the jetted AGNs across cosmic time~\cite{Ajello2009,Ghisellini2010,Volonteri2010,
Volonteri2011,Ghisellini2013,Sbarrato2015}. 

%-----------------------Section-3-Data Set------------------------------------------------------
\section{Data Set}\label{data}
In this study, we have used publicly available data in the literature. A brief description of the data set, obtained mainly from 3 sources, 
is given below:
\begin{itemize}
	\item{4LAC-DR3 Blazars Catalog:} The fourth catalog of AGNs detected by the \emph{Fermi}-LAT (4LAC), derived from the third data 
		release (DR3) of the 4FGL catalog based on 12 years of $\gamma$-ray data in the energy range above 50 MeV, provides measurements of 
		spectral parameters, SEDs, yearly light curves, and associations for all sources~\cite{Ajello2022,Abdollahi2022}. We consider 
		both low and high Galactic latitude sources of blazar type (i.e., BL Lacs, FSRQs, and Blazar Candidates of Uncertain types (BCUs)). Out of 
		3743 blazar-type sources, only 1811 have redshift measurements. Among 1811 blazars with known $z$, there are 875 BL Lacs, 792 FSRQs and 
		144 BCUs. 
	\item {Blazar Central Engines Catalog :} The central engine properties like  black hole mass (M$_{BH}$) and accretion disk luminosity 
		(L$_{disk}$) have been reported in literature \cite{Paliya2021} for 674 emission line blazars and 346 absorption line blazars observed 
		with the \emph{Fermi}-LAT. For these 1030 blazars, the values of M$_{BH}$ and L$_{disk}$ are derived using optical spectroscopic line features 
		reported either in the 16th data release of the Sloan Digital Sky Survey (SDSS-DR16)~\cite{Ahumada2020}) or from the NASA/IPAC Extragalactic 
		Database (NED)
 (\url{https://ned.ipac.caltech.edu/}) and SIMBAD Astronomical Databases 
 (\url{http://simbad.cfa.harvard.edu/simbad/}).
		It is important to note that since M$_{BH}$ values are derived from the scaling relations with line features, high correlations
		are expected between M$_{BH}$ and L$_{disk}$. In our present study, we have ignored 4 blazars whose L$_{disk}$ values are unusually low 
		($< 10^{10}$ erg~s$^{-1}$).

	\item {Blazar Emission Regions Catalog :} Recently, Fan et al. (2023)~\cite{Fan2023} have studied the properties of emission zones using the SED 
		fitting results of more than 2700 blazars reported in~\cite{Yang2023}. This large sample contains 1791 sources with known $z$ (750 FSRQs, 
		843 BL Lacs and 198 BCUs). The catalog reports derived values of the magnetic field ($B$) in the emission region, energy density of electrons 
		(U$_e$) and magnetic field energy density (U$_B$) from the SED fitting. For such blazars, we have collected the luminosities of $\gamma$-ray 
		at 1 GeV (log L$_\gamma$), X-ray at 1 keV (log L$_X$), optical at 2.43 $\times10^{14}$ Hz (log L$_O$) and radio at 1.4 GHz (log L$_R$) 
		from~\cite{Yang2022}. 
\end{itemize}

The blazar central engines catalog is a subset of the blazar emission regions catalog. And, both catalogs are subsets of the 4LAC-DR3 blazars catalog 
as they only include 4LAC-DR3 catalog sources with known redshift. Therefore, blazar central engines catalog (1026 sources) $\subseteq$ blazar emission regions 
catalog (1791 sources) $\subseteq$ 4LAC-DR3 blazars catalog (1811 sources).

%----------------------------------------------------Figure-1-Redshift-Distribution-----------------------------------------------------

%---------------------------------------Section-4-Resulst and Discussion--------------------------------------
\section{Results and Discussion}\label{results}

\subsection{Redshift Distribution}\label{sec:z}
\begin{figure}[H]
	\includegraphics[width=0.49\linewidth]{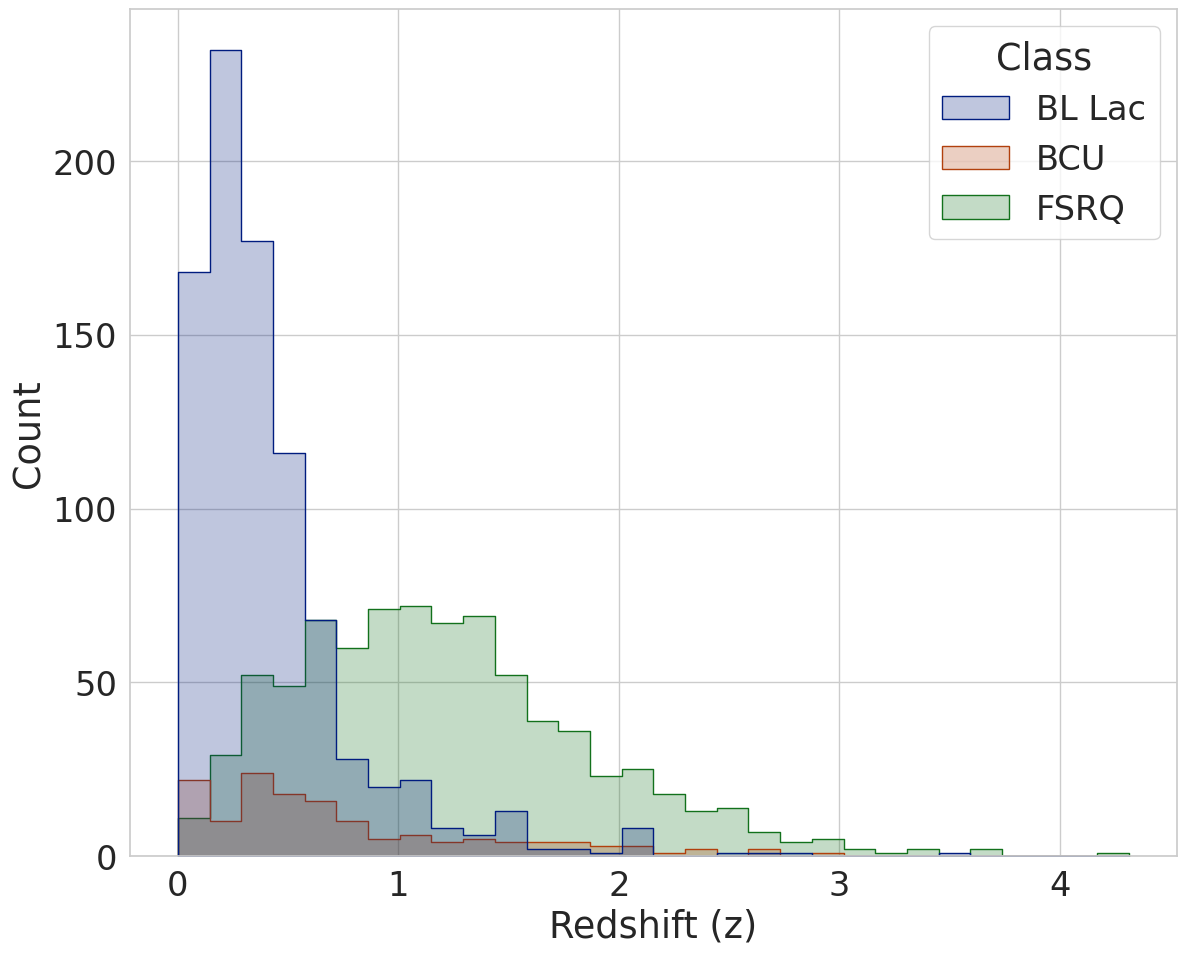}
	\includegraphics[width=0.49\linewidth]{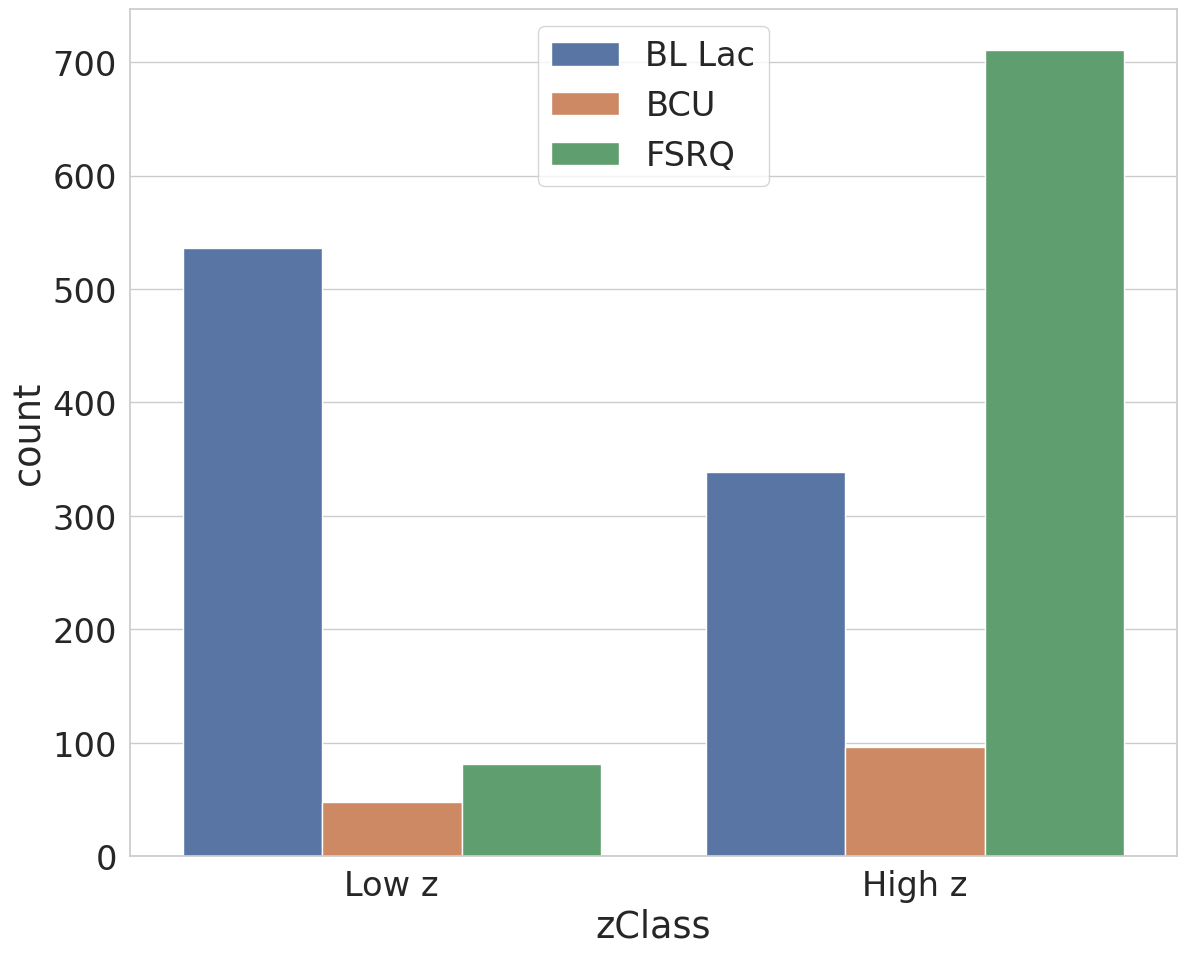}
\caption{(Left) Histograms of the redshifts for FSRQs, BL Lacs and BCUs taken from 4LAC-DR3 catalog~\cite{Ajello2022,Abdollahi2022}.
	 (Right) The number distributions of FSRQs, BL Lacs and BCUs in the high redshift ($ z > 0.4$) and low redshift ($ z \le 0.4 $) regimes.}
\label{fig:z_dist}
\end{figure}

The measured z-distribution of blazars listed in the 4LAC-DR3 blazars catalog is shown in the left panel of Figure~\ref{fig:z_dist}. 
The two blazar-subclasses, BL Lacs and FSRQs, show different $z$-distributions. The distribution of BL Lacs typically peaks around a median 
redshift value of $z\approx 0.31$ and is relatively narrow with a standard deviation of $\sigma \approx 0.39$. Whereas, the distribution of FSRQs 
peaks around a median of $z \approx 1.11$ and is broader with $\sigma=0.66$. The BCUs with relatively flatter distribution occupy the space between 
BL Lacs and FSRQs. The $z$-values play an important role in the classification of BCUs in the two subclasses of blazars~\cite{Tolamatti2023}. 
The right panel of Figure~\ref{fig:z_dist} shows the populations of BL Lacs, FSRQs, and BCUs in different redshift regions. It is observed that 
in the low-z region, the blazar population is mostly dominated by BL Lacs ($\sim 80$\%) and few FSRQs (13\%) and BCUs (7\%). In the high-z regime, 
FSRQs dominate the blazar population with 62\% contribution whereas BL Lacs (30\%) and BCUs (8\%) correspond to the remaining 38\% population. 
Thus, BL Lacs are typically located in the nearby universe, within a physical distance of about 2--3 billion light-years, whereas FSRQs are found 
at longer distances than BL Lacs, and may be associated with more distant and younger galaxies. The exact cause of these redshift 
distributions is not well understood, yet many explanations have been proposed in the literature. The first explanation is the selection bias 
in observations, i.e., BL Lacs are low luminous sources and it is very difficult to detect them at high redshifts. Even if they are detected, 
measurement of their redshift is challenging due to their line-less optical spectra~\cite{Abdo2009,Giommi2012}. Thus, the population 
of BL Lacs is less numerous at higher redshifts. As only strongly luminous sources can be detected in the high-z regime, FSRQs being intrinsically more luminous 
than BL Lacs dominate the high redshift Universe. The second explanation is that FSRQs evolve into BL Lacs over cosmological timescales. Therefore, 
high-z sources (early and young sources) are FSRQs and they evolve into BL Lacs, which are found at lower 
redshifts~\cite{Bottcher2002,Ghisellini2011,Ajello2014}. The highest redshift BL Lacs and FSRQ, detected by the \emph{Fermi}-LAT, are 
4FGL J1219.0+3653 and 4FGL J1510.1+5702 (with radio association of NVSS J151002+570243) located at $z~\approx$ 3.59 and z$\approx$ 4.31, 
respectively~\cite{Paliya2020b,Ackermann2017}. They correspond to a population of blazars existing just a few billion years after the Big Bang.

%--------------------------------------Figure-2-----------------------------------------------------------

%--------------------------------------------------------------------------------
\subsection{$\gamma$-ray Spectral Features} \label{sec:LAT_Spectrum}
\begin{figure}[H]
	\includegraphics[width=0.48\linewidth]{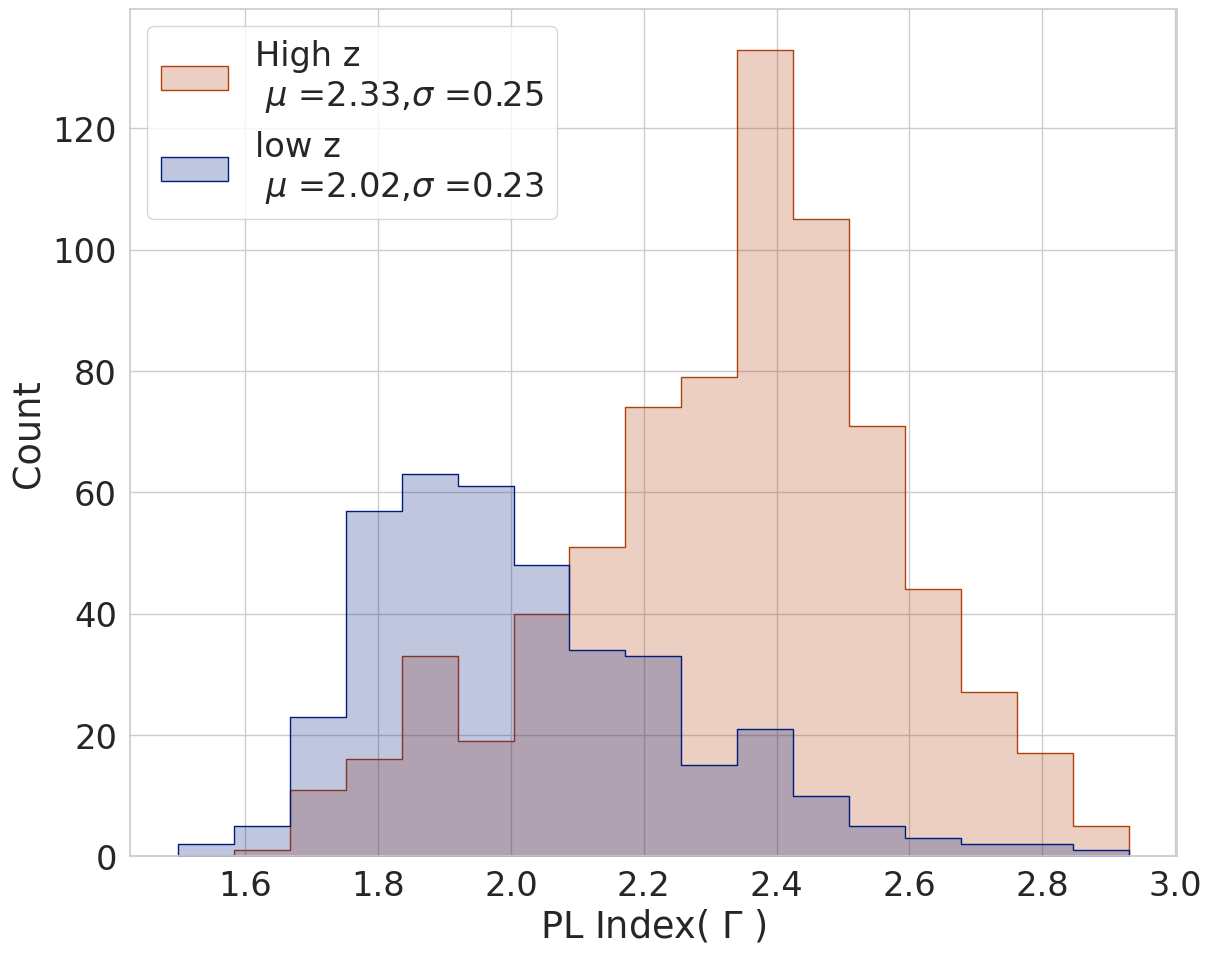}
	\includegraphics[width=0.48\linewidth]{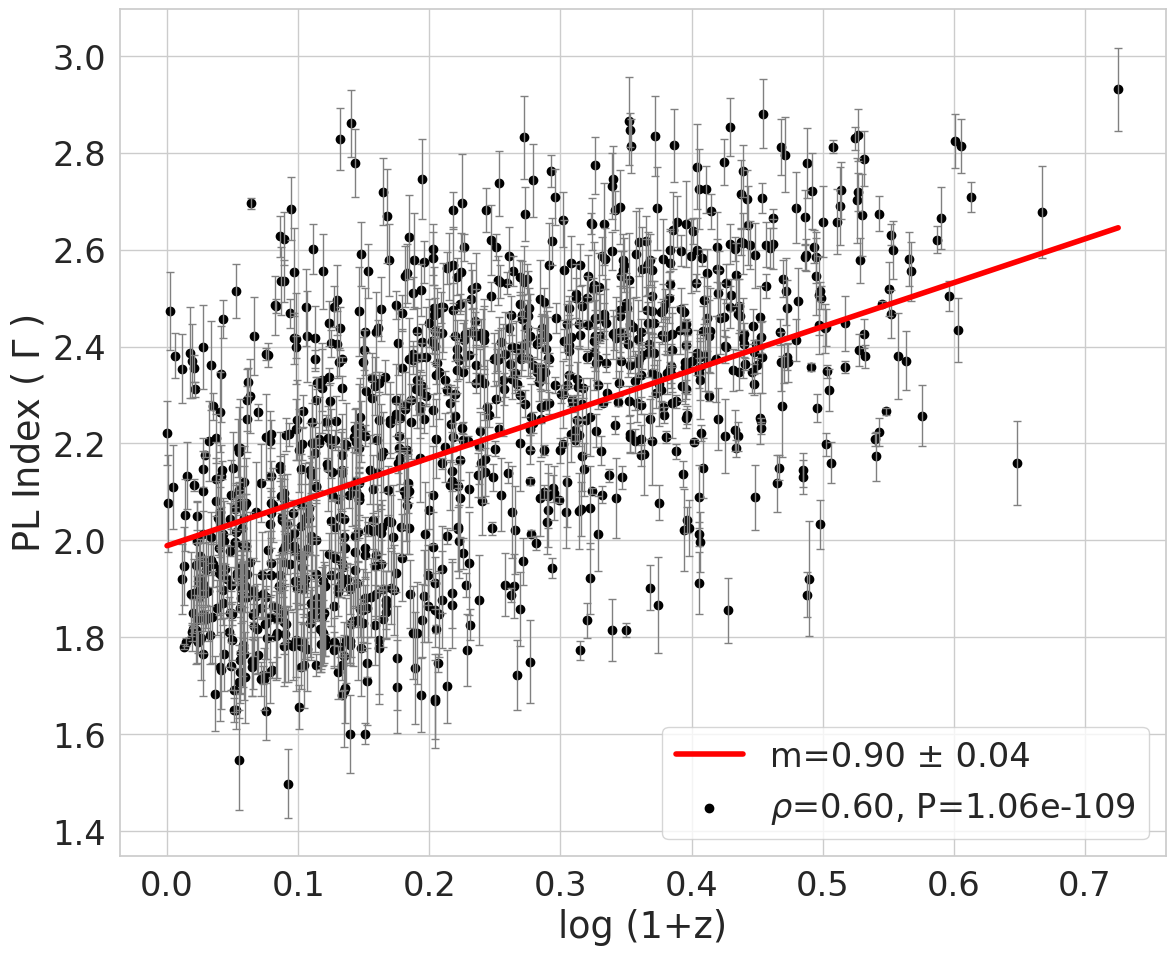}
	\includegraphics[width=0.48\linewidth]{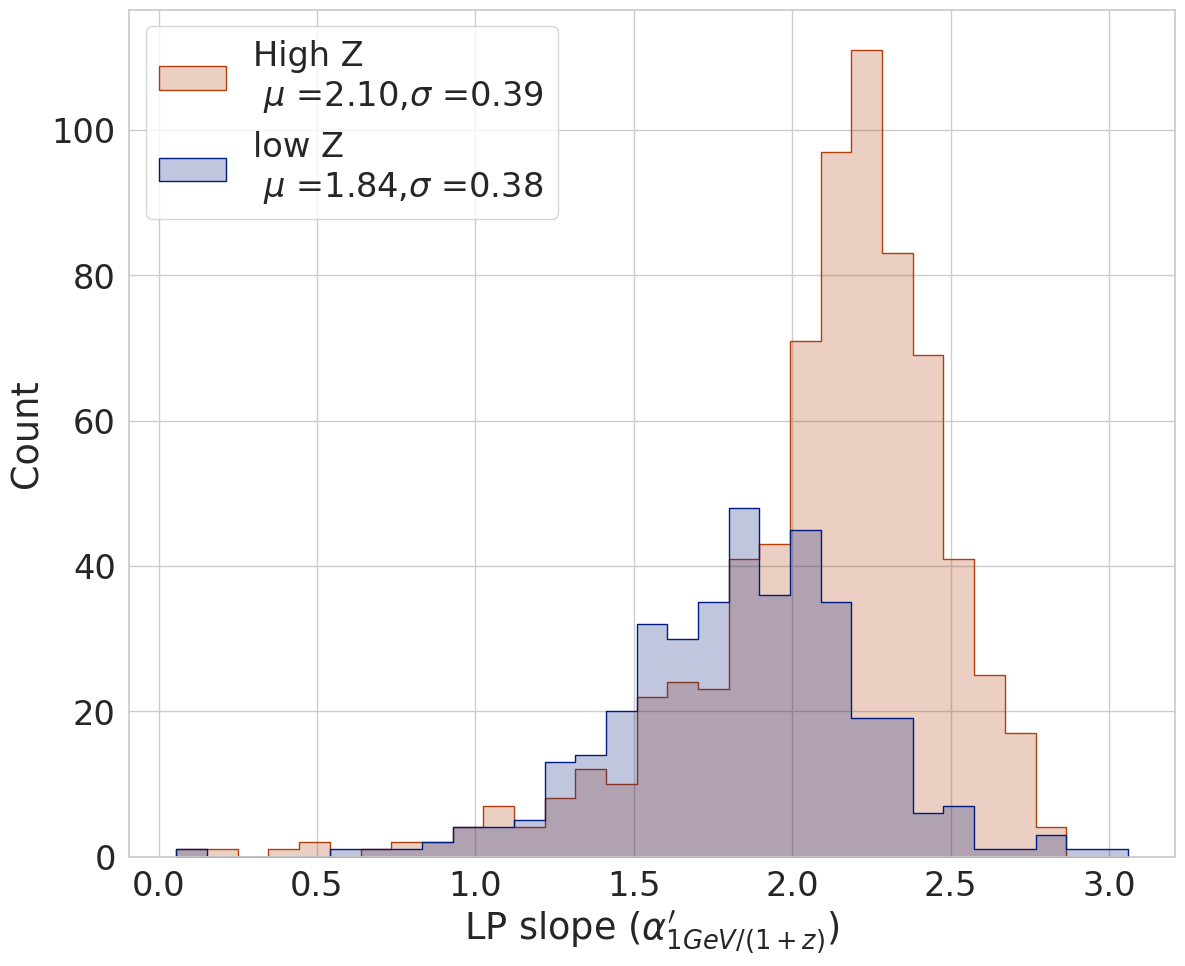}
	\includegraphics[width=0.48\linewidth]{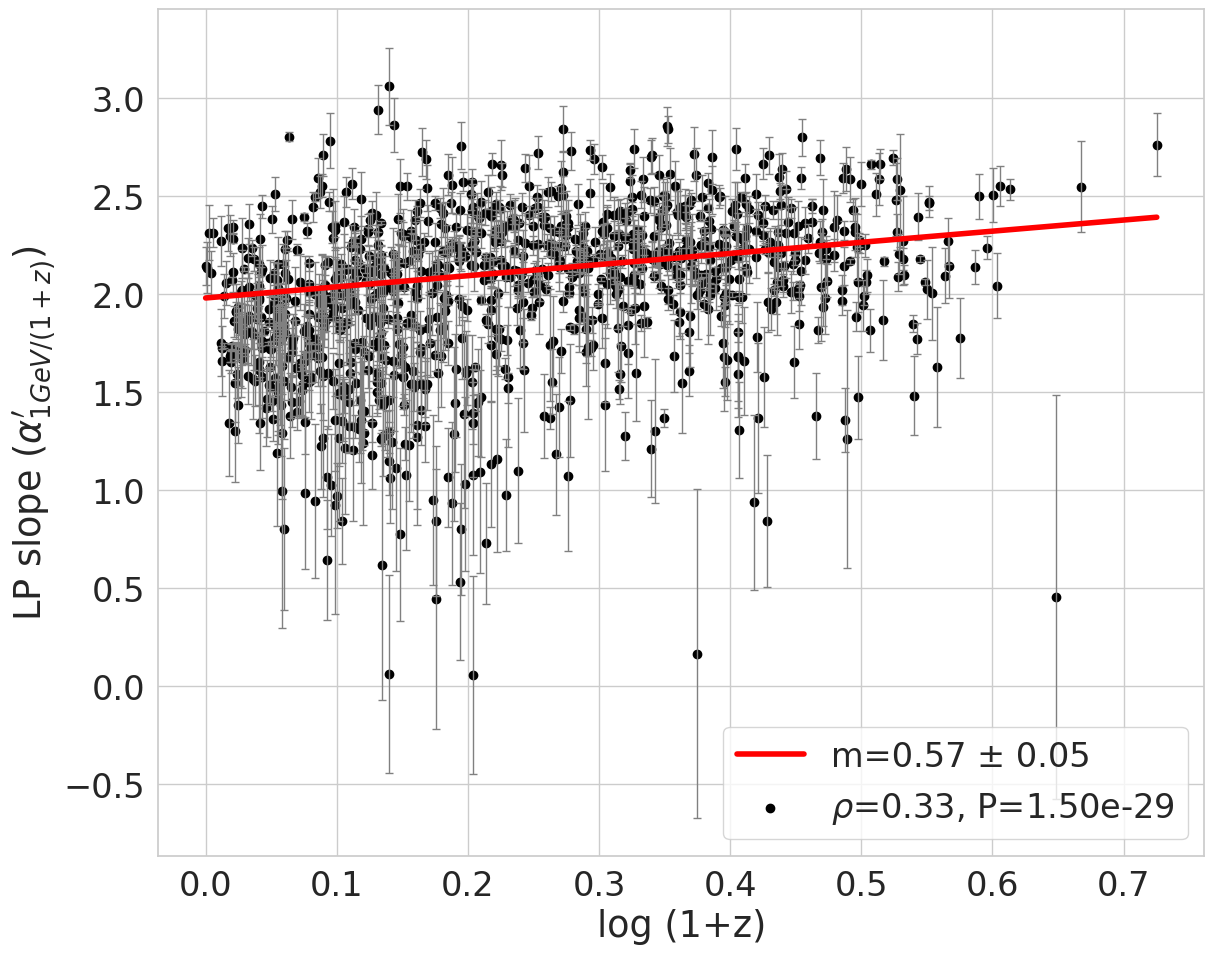}
	\includegraphics[width=0.48\linewidth]{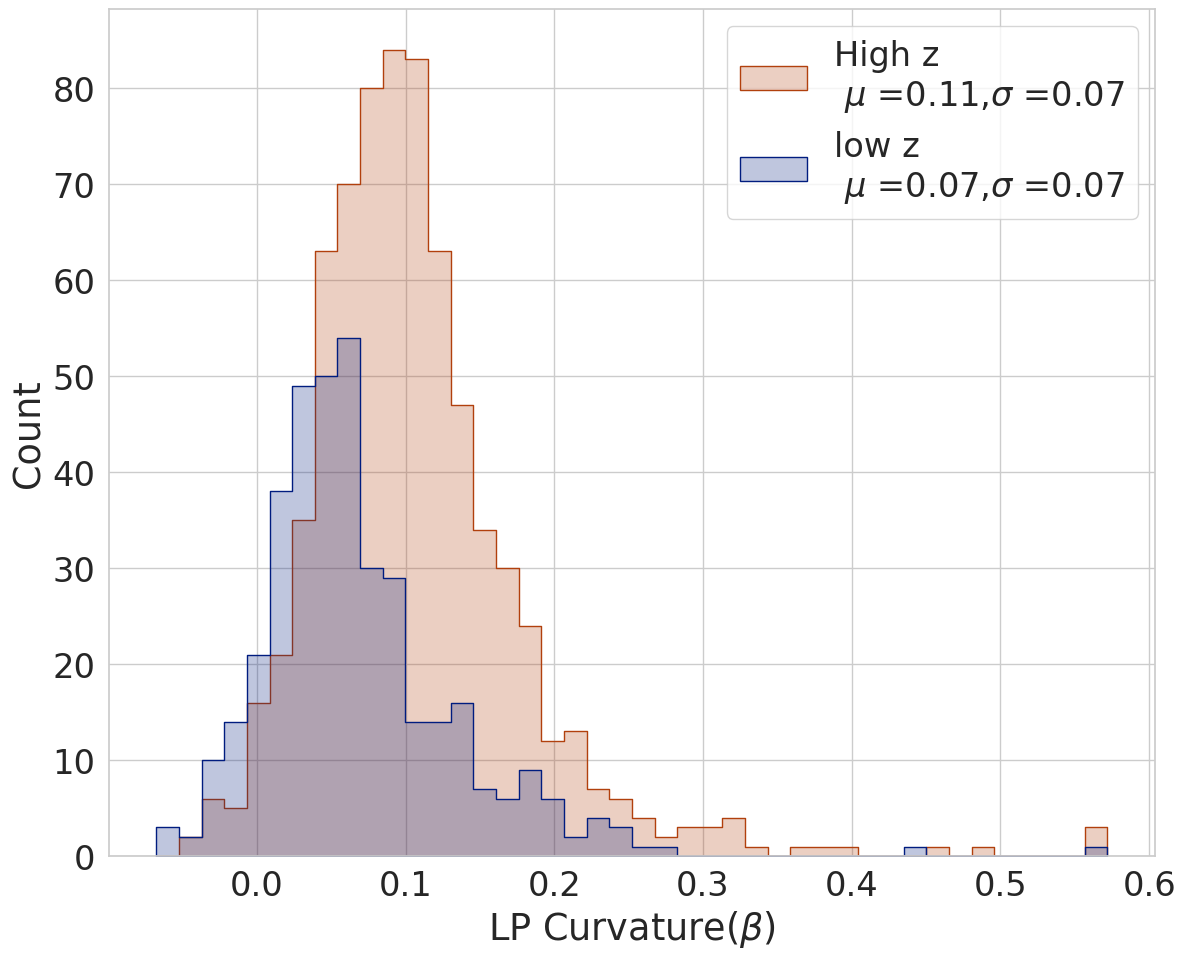}
	\includegraphics[width=0.48\linewidth]{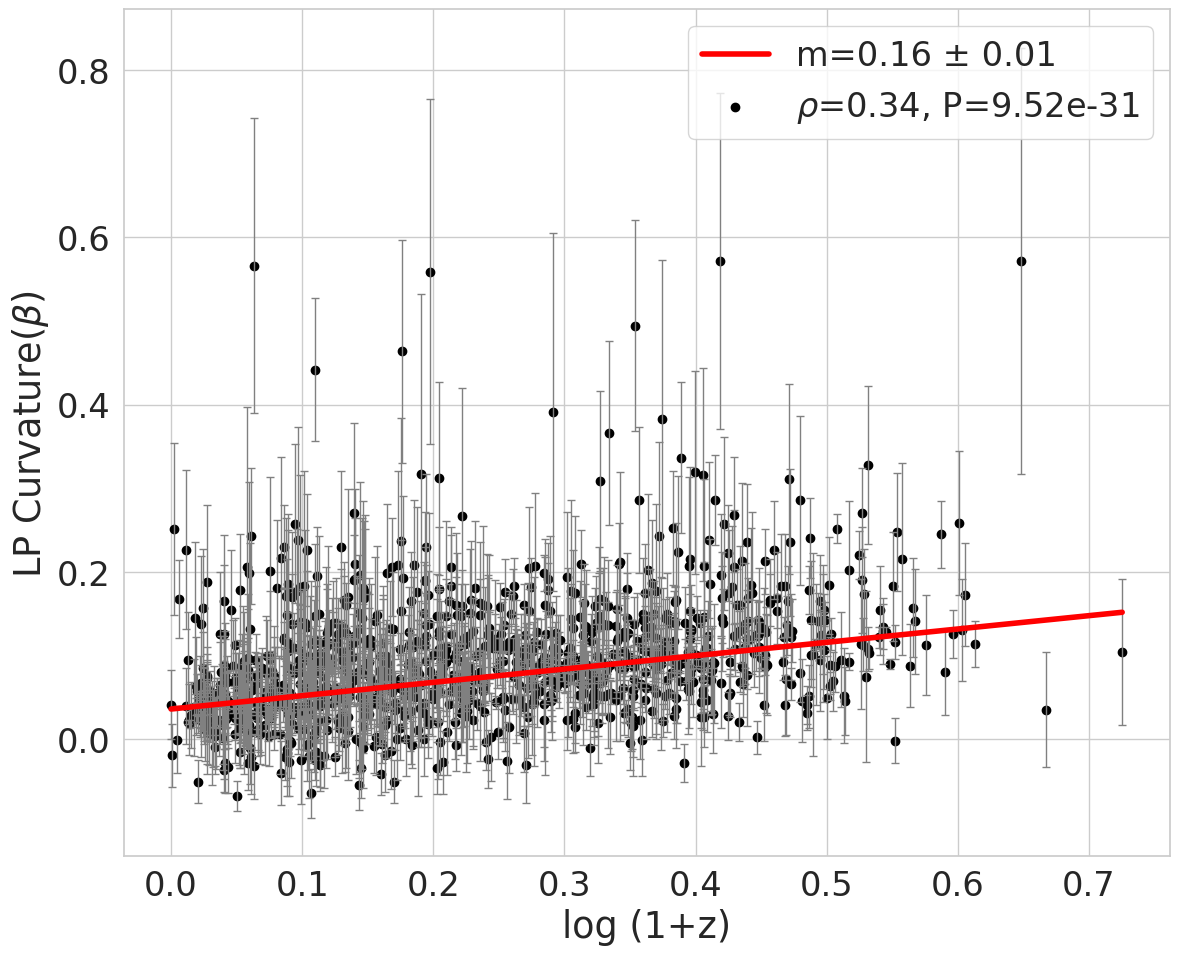}
	\caption{(Left) Histograms of the Power-Law (PL) spectral index ($\Gamma$),  Log Parabola (LP) index ($\alpha^\prime$) and
		 curvature parameter ($\beta$) of blazars measured by the \emph{Fermi}-LAT. (Right)  Variations of $\Gamma$, $\alpha^\prime$ 
		 and $\beta$ parameters as a function of redshift (log (1 + z)).}
\label{fig:LAT_Spectrum}
\end{figure}

For the spectral study, we consider sources reported in the 4LAC-DR3 blazars catalog with a detection significance 
(\emph{'Signif\_Avg'} in the Fermi catalog) of more than 10$\sigma$ in the energy range 100 MeV--1 TeV. There are 1111 
such sources comprising 535 BL Lacs, 521 FSRQs, and 55 BCUs. In this catalog, the observed $\gamma$-ray flux 
points (i.e., without taking EBL attenuation into account) have been fitted with both Power-Law (PL) 
\begin{equation}
	\frac{dN}{dE}= K {\Bigl(\frac{E}{Eo}\Bigr)}^{-\Gamma}
\end{equation}
and Log Parabola (LP) 
\begin{equation}
	\frac{dN}{dE}= K {\Bigl(\frac{E}{Eo}\Bigr)}^{-\alpha -\beta \text{~ln}(E/Eo)}
\end{equation}
spectral models. Here, $K$ is the flux density (erg~cm$^{-2}$~s$^{-1}$), $Eo$ is the pivot energy 
(in MeV) at which the error in differential flux is minimal, $\Gamma $ is PL spectral index, $\alpha$ is LP index and 
$\beta$ is the LP curvature parameter. For a fair comparison, we have calculated the modified LP index ($\alpha^{\prime}$) by computing 
the slope of the LP spectrum at energy 1 GeV/(1 + z). This energy corresponds to the photon energy of 1 GeV in the source rest frame. 
Therefore, $\alpha^\prime$ represents the slope of the observed spectrum at the energy corresponding to 1 GeV in the source frame.
Left panels in Figure~\ref{fig:LAT_Spectrum} show distributions of $\Gamma$, $\alpha^\prime$ and $\beta$ for high and low-z blazars. 
\textls[-15]{The two sample Kolmogorov--Smirnov (K-S) test results in \emph{p}-values of  $\approx$0 for all three parameters. This 
suggests that distributions of $\Gamma$, $\alpha^\prime$, and $\beta$ parameters are intrinsically different for high and low-z blazars. 
A Gaussian fit to these distributions gives mean and standard deviation, $\mu \pm \sigma$, values of 2.33 $\pm$ 0.25 (2.02 $\pm$ 0.23), 
2.10 $\pm$ 0.39 (1.84 $\pm$ 0.38), and 0.11 $\pm$ 0.07 (0.07 $\pm$ 0.7) for $\Gamma$, $\alpha^\prime$ and $\beta$, respectively, in case 
of high-z (low-z) blazars. The right panels in Figure~\ref{fig:LAT_Spectrum} depict variations of these parameters as a function of (log (1 + z)).}
Pearson correlation analysis shows that $\Gamma$, $\alpha^\prime$ and $\beta$ parameters are positively correlated with (log (1 + z)) 
with the correlation coefficient ($\rho$) values of 0.6, 0.33, and 0.34, respectively, and \emph{p}-values of $\approx$0.
Therefore, blazars at higher redshifts will have softer $\gamma$-ray spectra (higher values of $\Gamma$ and $\alpha^\prime$) and 
with more curvature (higher values of $\beta$) than low-z blazars. Similar correlations have also been obtained between BL Lacs 
and FSRQs in the literature~\cite{Iyida2022,Ajello2022}. This may be the manifestation of the underlying particle spectrum involved 
in the HE emission (detail discussion in Section~\ref{sec:emission_zone}) and absorption of HE $\gamma$-ray photons by EBL via 
$\gamma$-$\gamma$ pair production during their propagation from source to the observer~\cite{Gould1966,Gould1967}. The absorption probability 
of HE photons traveling over cosmological distances is characterized by optical depth ($\tau$) or opacity, which strongly depends on $z$, $\gamma$-ray 
photon energy (E) and density of EBL photons~\cite{Stecker2006,Singh2014}. The EBL absorption effects start dominating at energies above 30 GeV 
for redshifts beyond $z~>~0.4$~\cite{Tolamatti2022a}. Therefore, the intrinsic (emitted) $\gamma$-ray spectrum of the source gets modified and 
the observed spectrum is given as 
\begin{equation}
	{\Bigl(\frac{dN}{dE}\Bigr)}_{obs} = {\Bigl(\frac{dN}{dE}\Bigr)}_{int} e^{-\tau (E,z)}.
\end{equation}
For high-z blazars, $\tau$ will be large due to strong EBL absorption. Therefore, the observed photon spectrum will be soft with more curvature 
\cite{Finke2010,Dominguez2011}. Hence, the detection and accurate measurements of HE and VHE spectra of high-z blazars are important (as they will have 
an imprint of EBL absorption effects) in constraining the EBL energy density in extragalactic space~\cite{Stecker2006,Mankuzhiyil2023}. 

%------------------------------Figure-3--------------------------------------------------

%------------------------------------------------------------
\subsection{Luminosity Features}\label{sec:Lumi}
\begin{figure}[H]
	\includegraphics[width=0.45\linewidth]{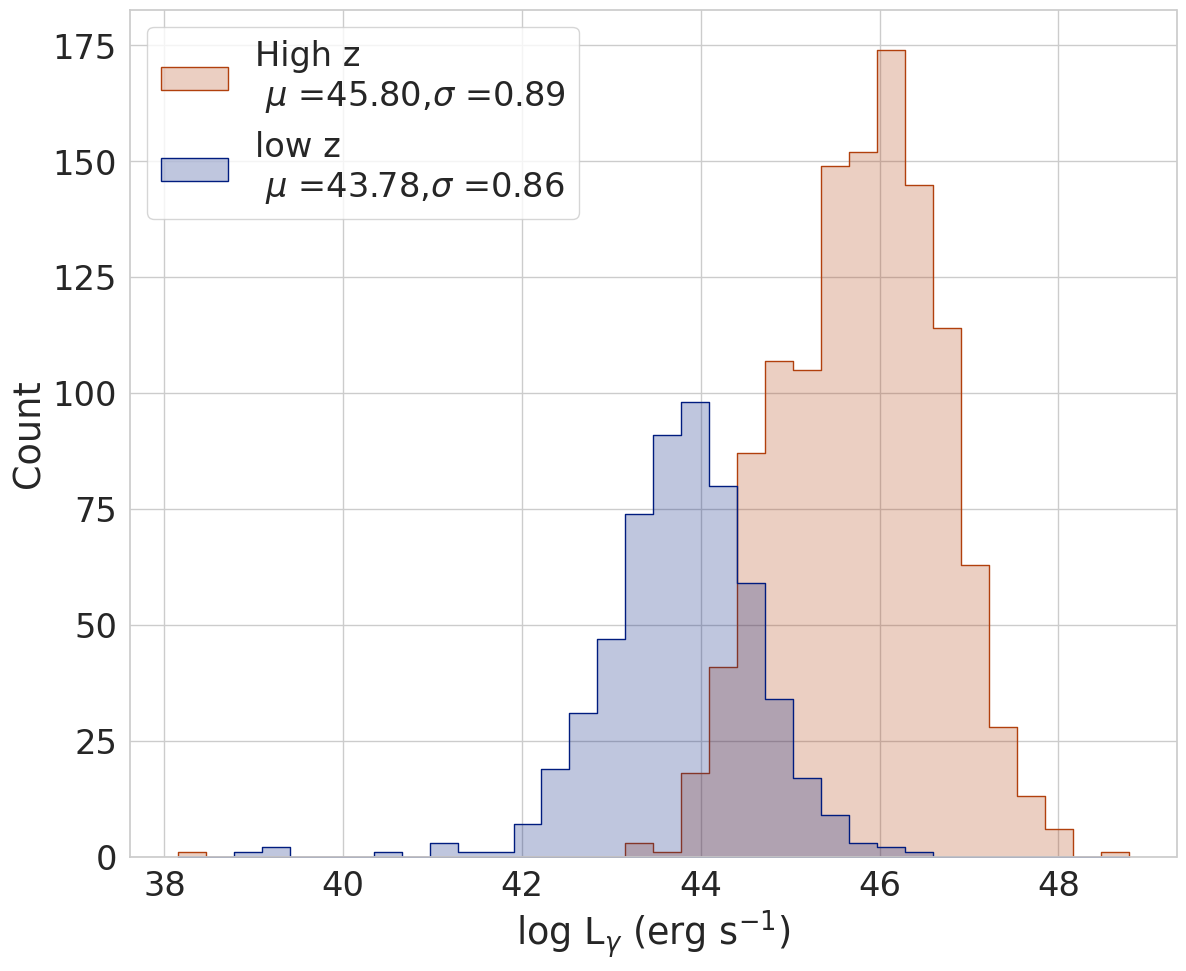}
	\includegraphics[width=0.45\linewidth]{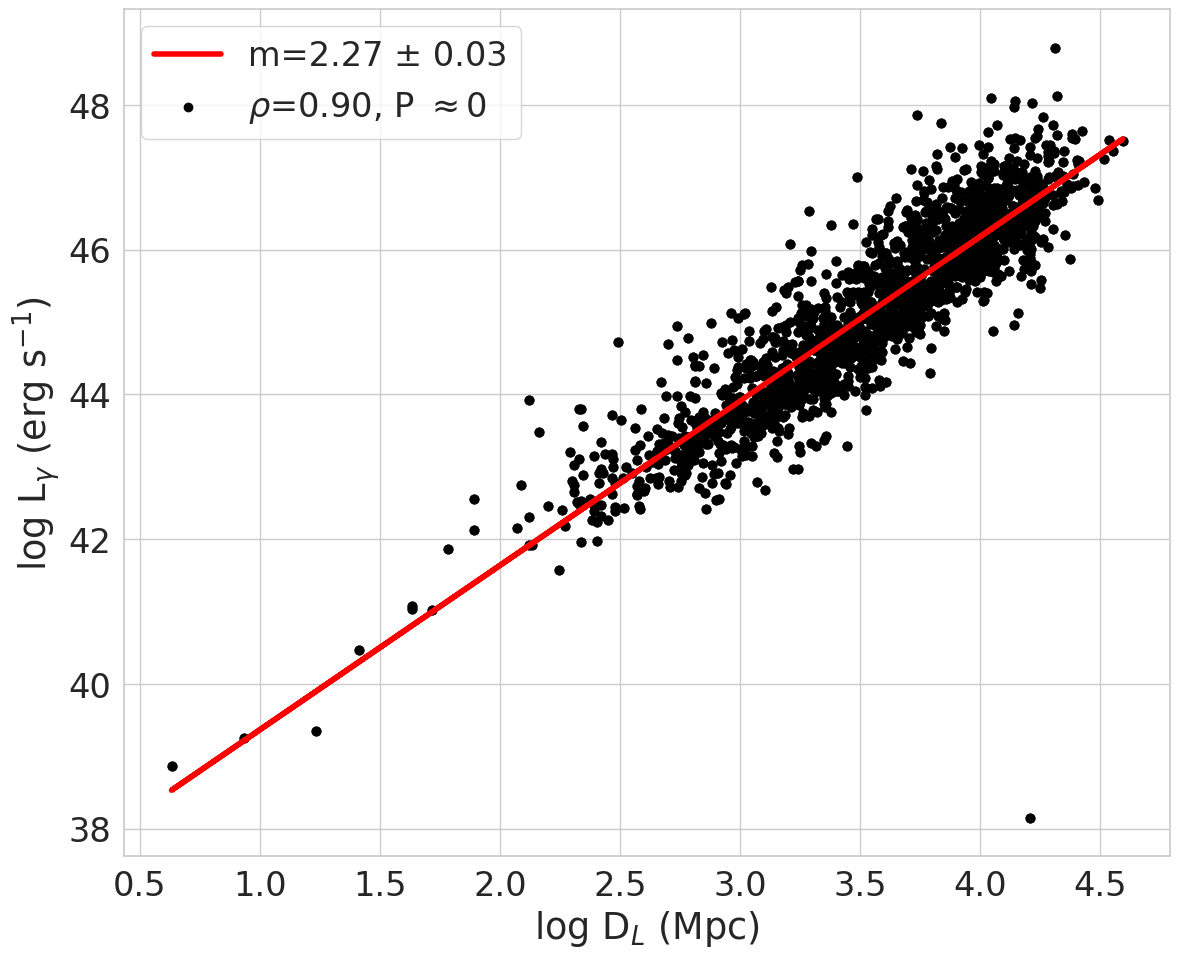}
	\includegraphics[width=0.45\linewidth]{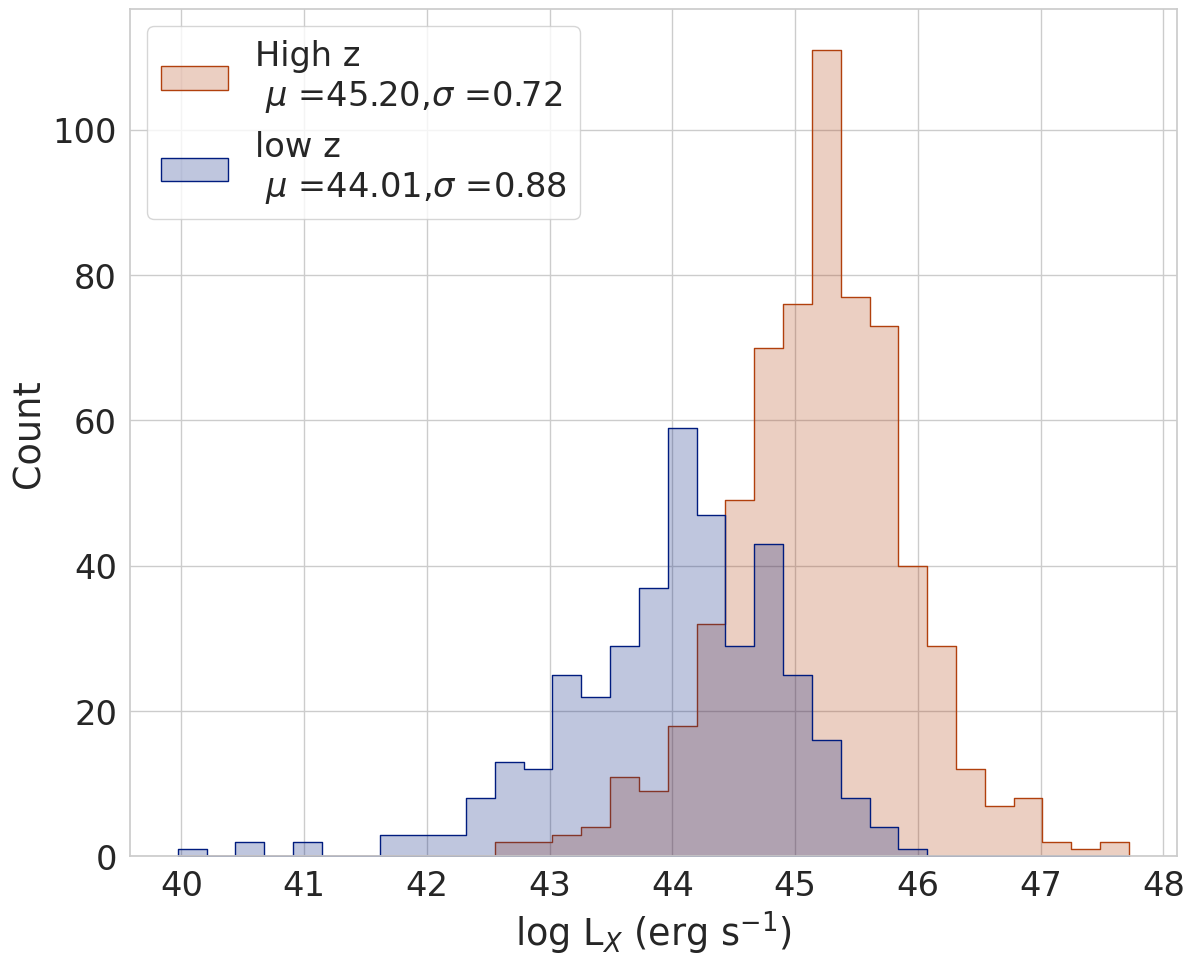}
	\includegraphics[width=0.45\linewidth]{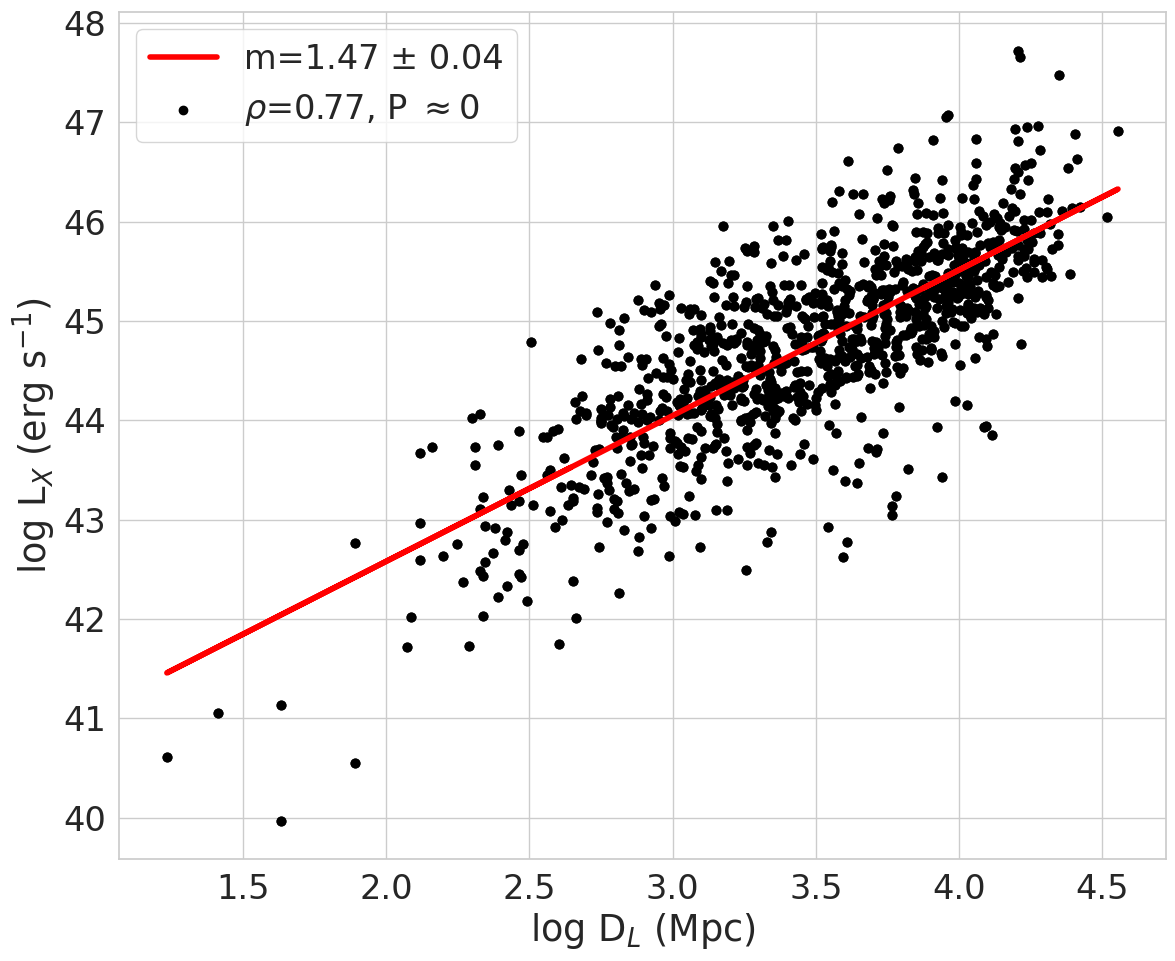}
	\includegraphics[width=0.45\linewidth]{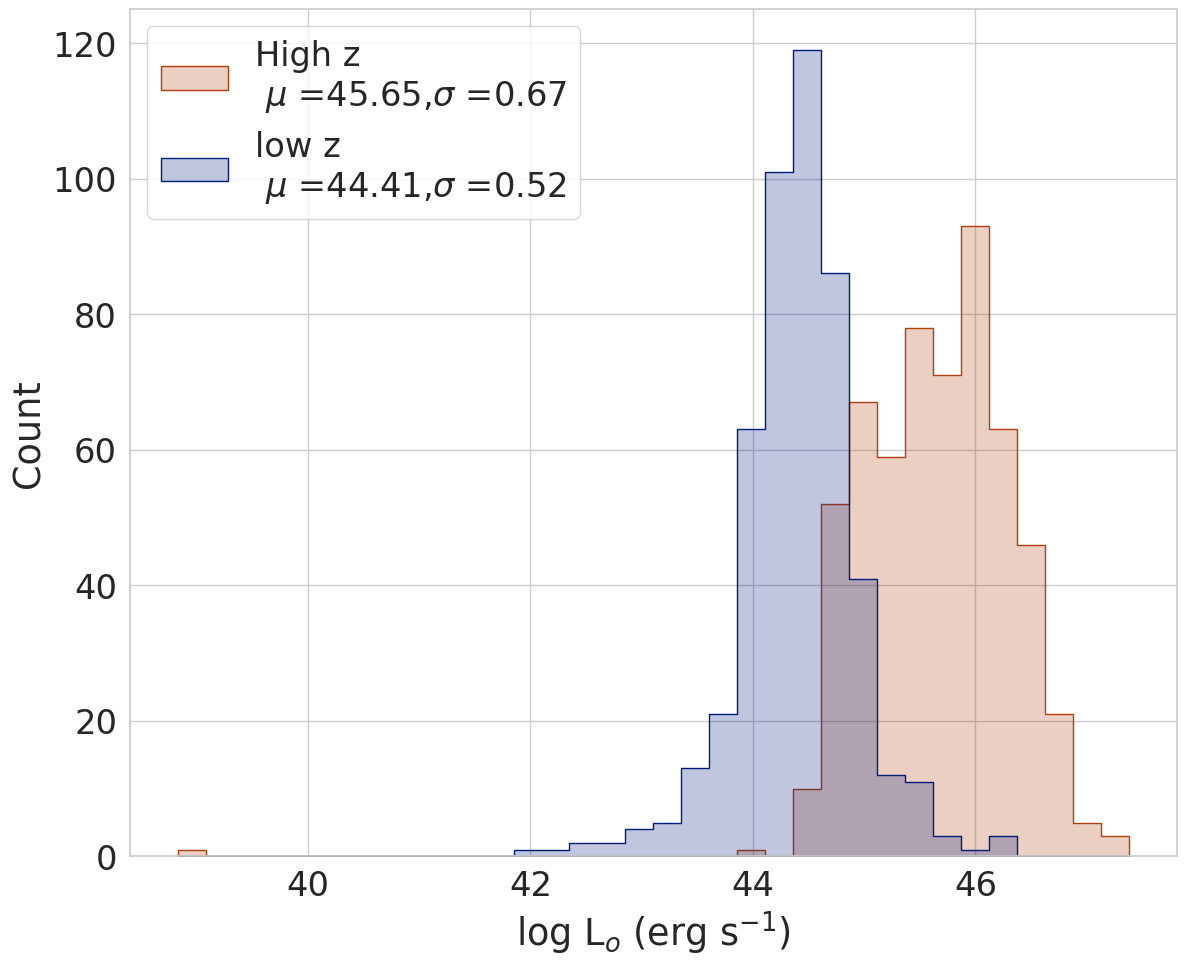}
	\includegraphics[width=0.45\linewidth]{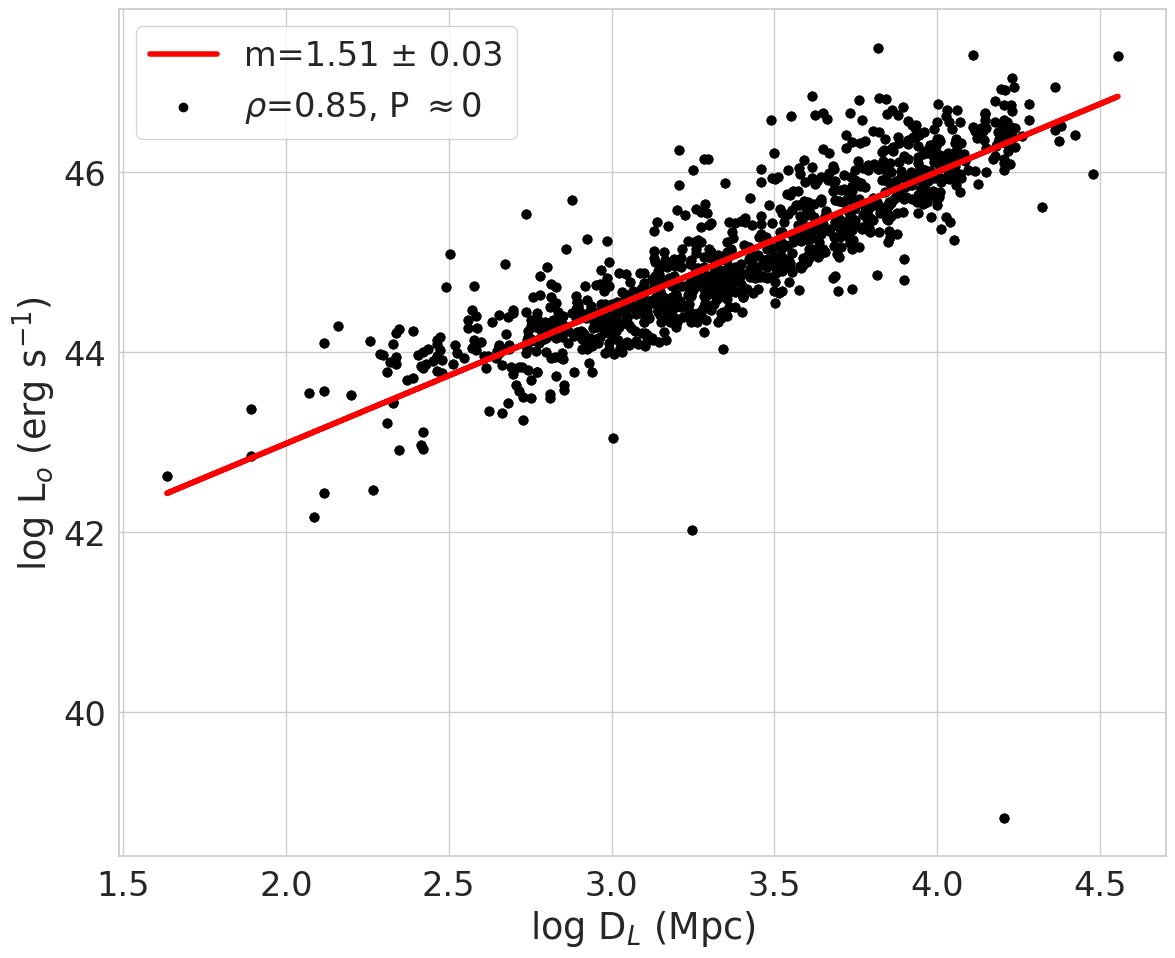}
	\includegraphics[width=0.45\linewidth]{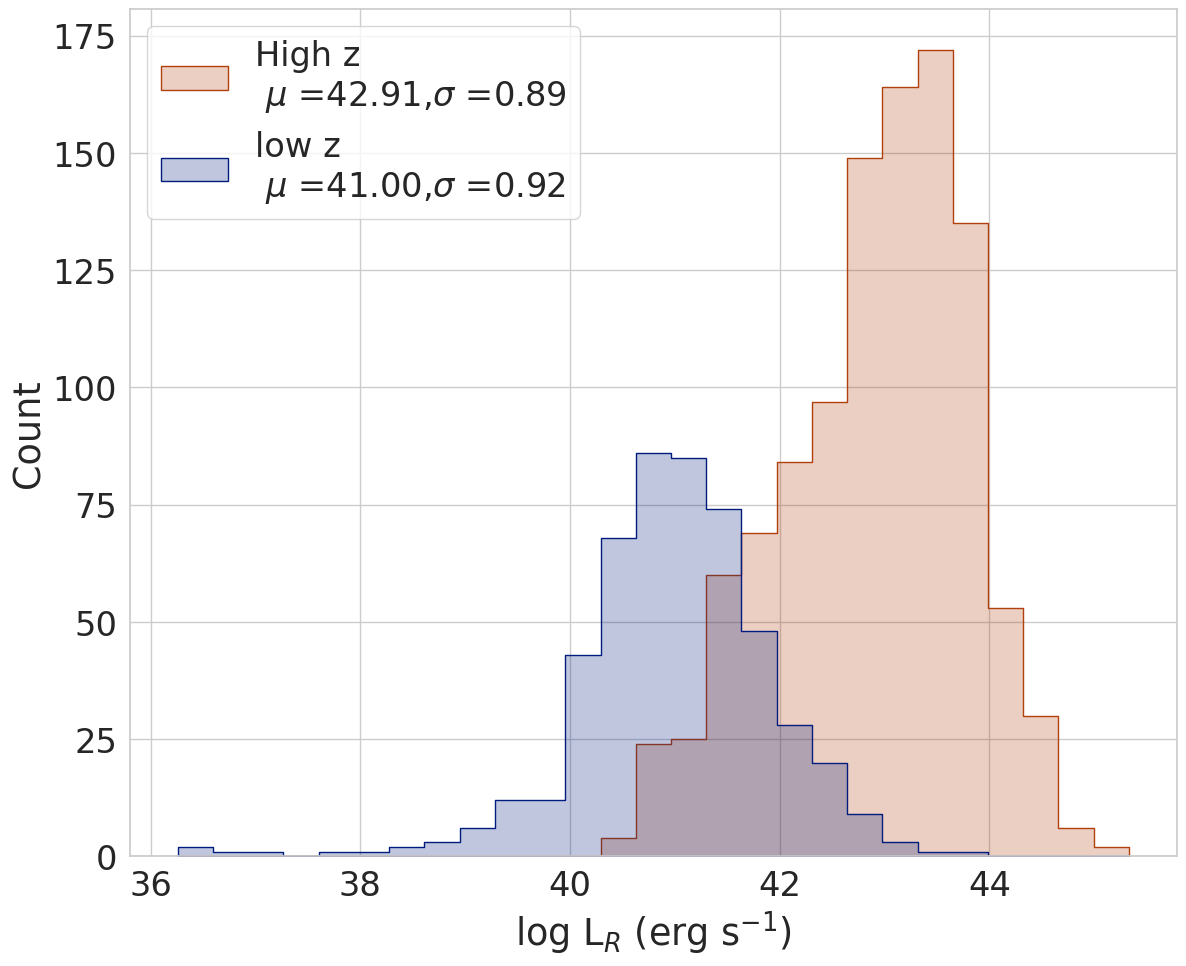}
	\hspace{1.3cm}\includegraphics[width=0.45\linewidth]{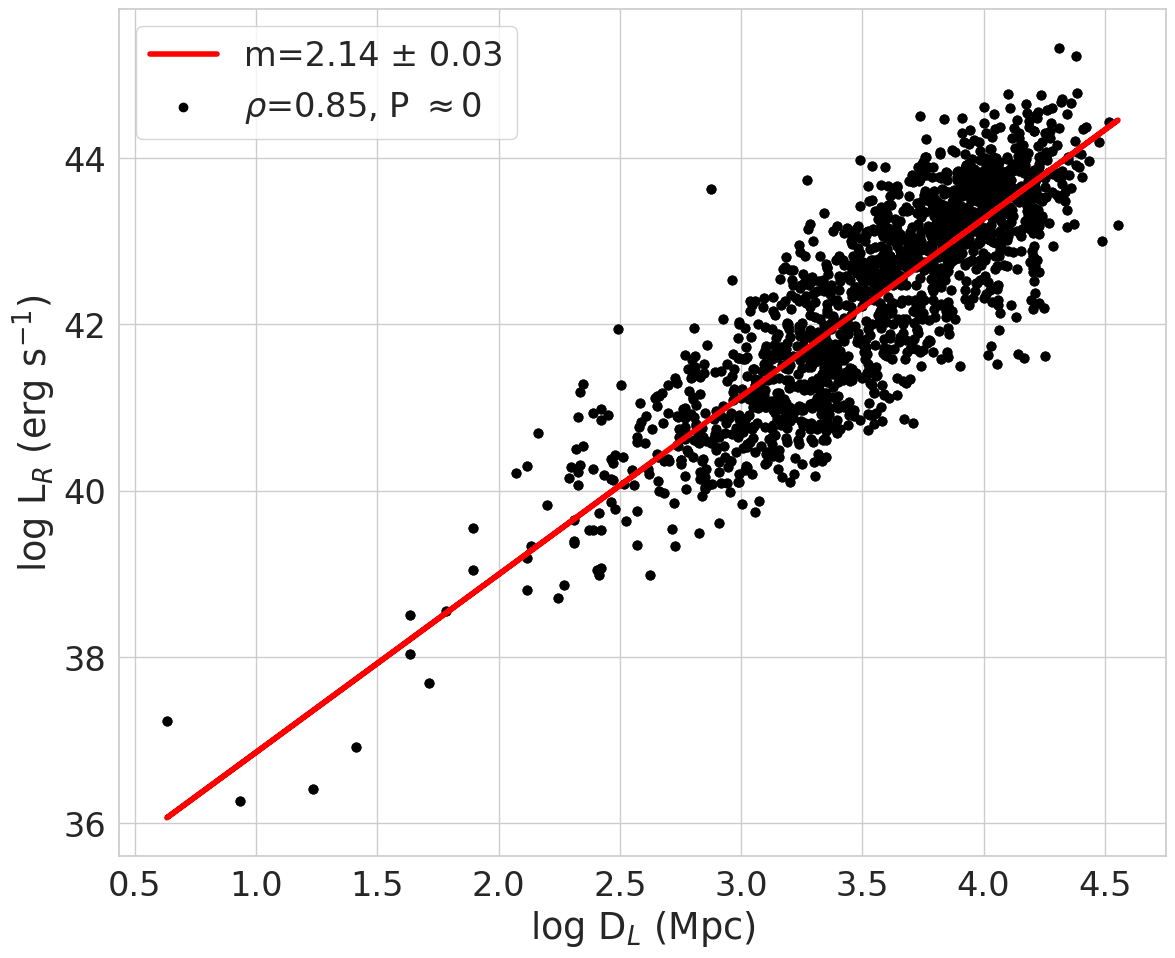}
	\caption{(Left) Histograms of the luminosities of $\gamma$-ray at 1 GeV (log L$_\gamma$), X-ray at 1 keV (log L$_X$), 
		optical at 2.43 $\times 10^{14}$ Hz (log L$_o$) and radio at 1.4 GHz (log L$_R$) for sample high and low-z blazars.  
		(Right)  Variations of log L$_\gamma$, log L$_X$, log L$_o$ and log L$_R$ as a function of luminosity 
		distance (log $D_L$).}
	\label{fig:Luminosities}
\end{figure}

For luminosity studies, we have selected sources from the blazar emission regions catalog. Out of 1791 sources in the catalog, 
1789 have  $\gamma$-ray luminosities (L$_\gamma$) at 1 GeV, 1030 have X-ray luminosities (L$_X$) at 1 keV, 1059 have optical luminosities 
(L$_O$) at 2.43 $\times$ 10$^{14}$ Hz and 1581 have radio luminosities (L$_R$) at 1.4 GHz. The left panels in Figure~\ref{fig:Luminosities} 
show distributions of log L$_\gamma$, log L$_X$, log L$_O$ and log L$_R$ for high and low-z blazars~\cite{Yang2022}. The right panels represent 
variations of these parameters as a function of luminosity distance (D$_L$) measured in Mega-parsec (Mpc). Note that, D$_L$ is a function of 
z. A Gaussian fit to the distributions of log L$_\gamma$, log L$_X$,  log L$_O$, and log L$_R$ gives $\mu \pm \sigma$ values of 45.80 $\pm$ 0.89 
(43.78 $\pm$ 0.86), 45.20 $\pm$ 0.72 (44.01 $\pm$ 0.88),  45.65 $\pm$ 0.67 (44.41 $\pm$ 0.52), and 42.91$\pm$ 0.89 (41.0 $\pm$ 0.92)
for high-z (low-z) blazars, respectively. We observe that, in all the wavebands considered above, high-z blazars are at least an order of magnitude more 
luminous than low-z blazars. All the luminosities log L$_\gamma$, log L$_X$, log L$_O$, and log L$_R$ show strong positive correlations with log D$_L$ 
having corresponding $\rho$ values of 0.90, 0.77, 0.85, and 0.85, respectively. We also perform a linear regression analysis of the scatter plots 
of these luminosities vs log D$_L$. The slopes of linear regression indicate that observed L$_\gamma$, L$_X$,   L$_O$ and  L$_R$ vary as 
$\sim$ D$_L^{2.27}$, D$_L^{1.47}$, D$_L^{1.51}$ and D$_L^{2.14}$, respectively. As mentioned in Section~\ref{sec:z}, these observed features 
are consistent with the fact that high-z sources with strong luminosity can only be detected due to selection bias. The high-z regime is dominated by 
FSRQs, which are intrinsically more luminous than BL Lacs, which are mostly observed at lower redshifts. Thus, high-z blazars (mostly FSRQs)  are 
the most luminous sources observed in the sky across all wavebands.

\subsection{Broadband SED Properties}\label{sec:SED}
\begin{figure}[H]
%	\centering
	\includegraphics[width=0.4\linewidth]{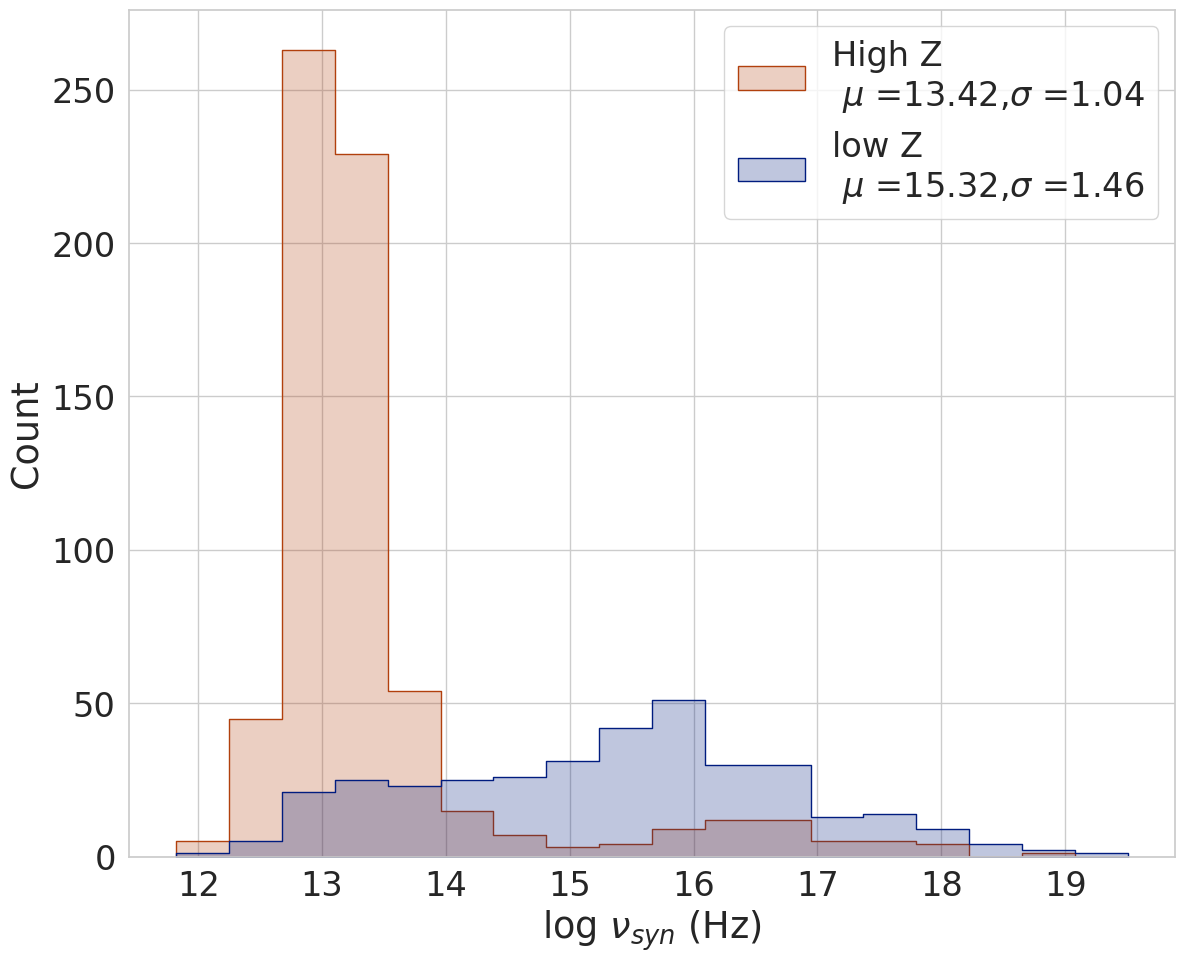}
	\includegraphics[width=0.4\linewidth]{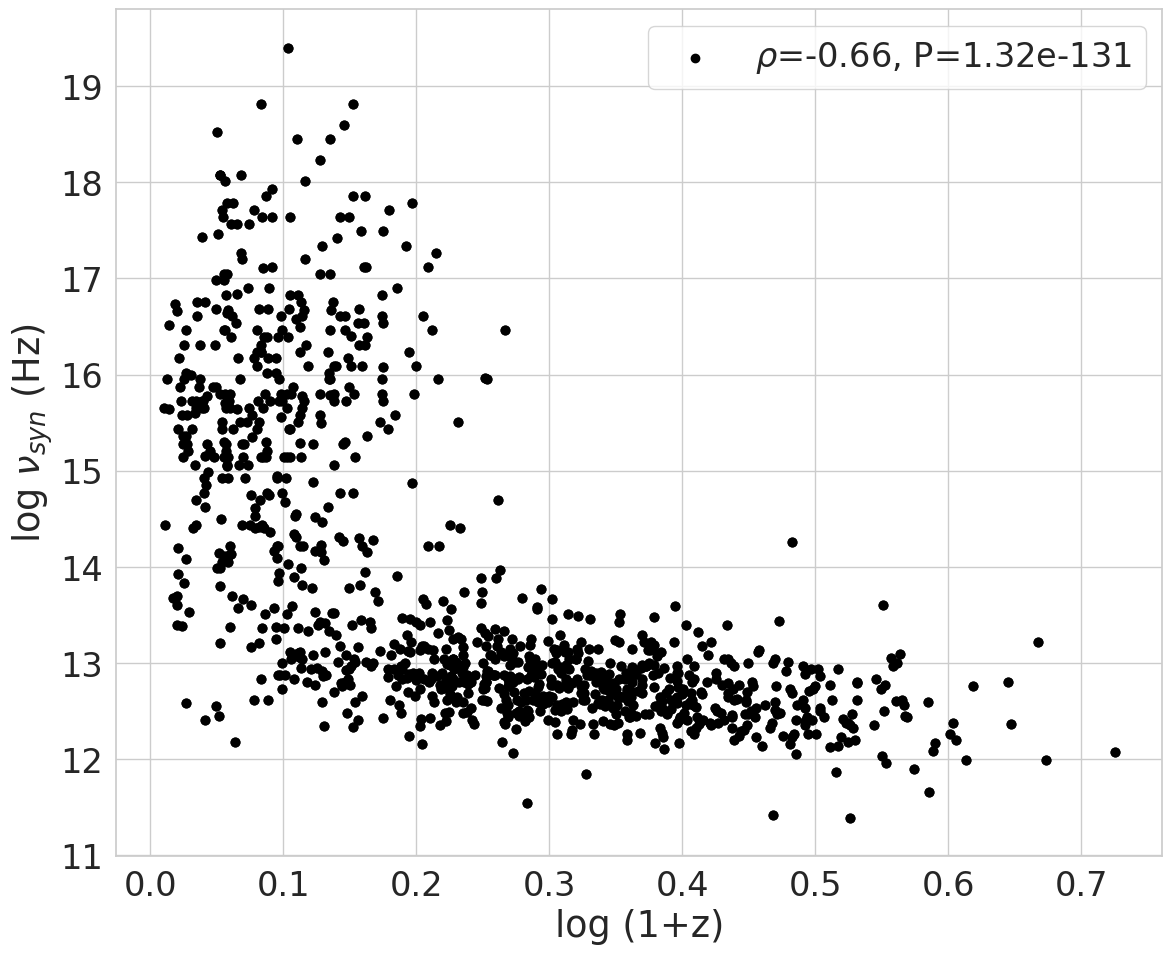}
	\includegraphics[width=0.4\linewidth]{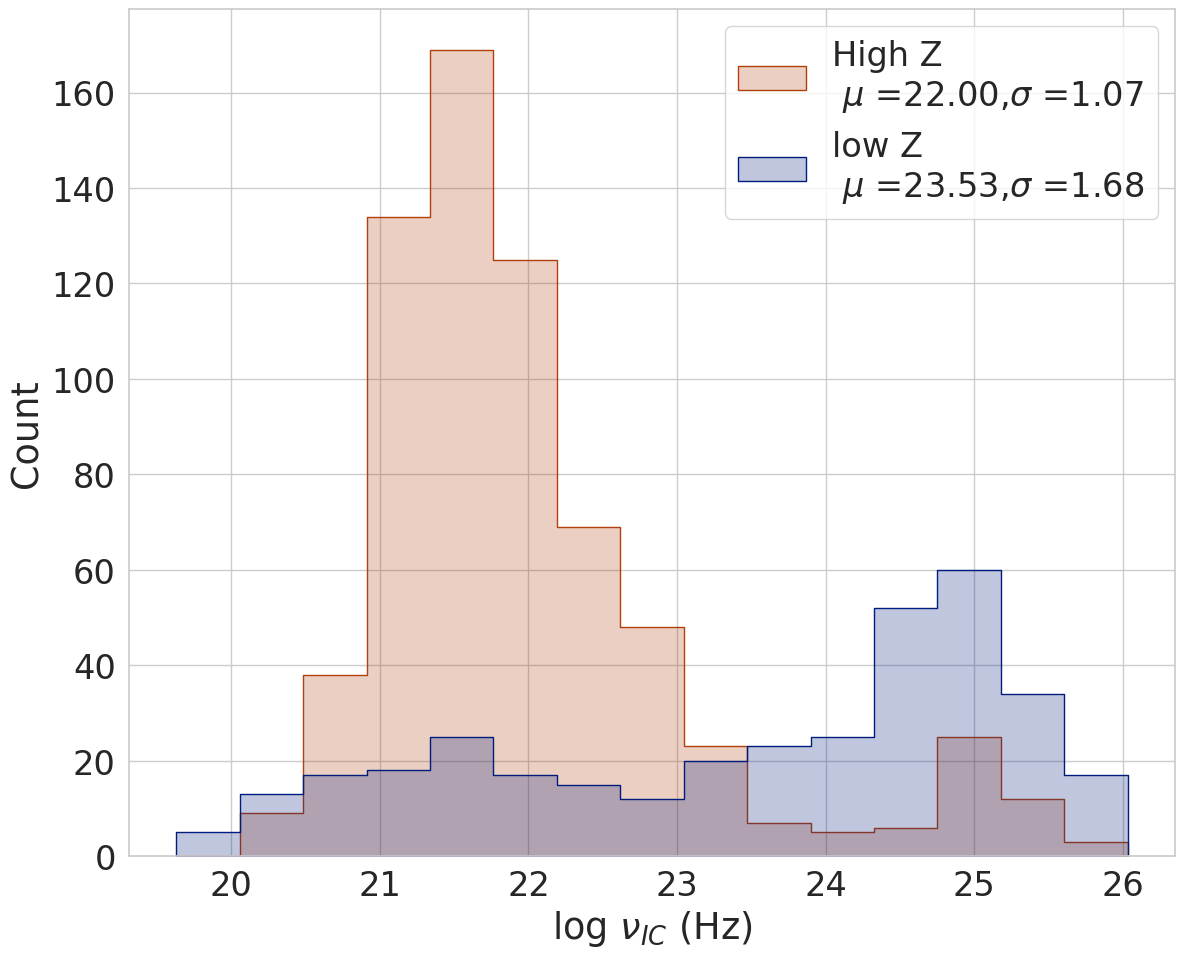}
	\includegraphics[width=0.4\linewidth]{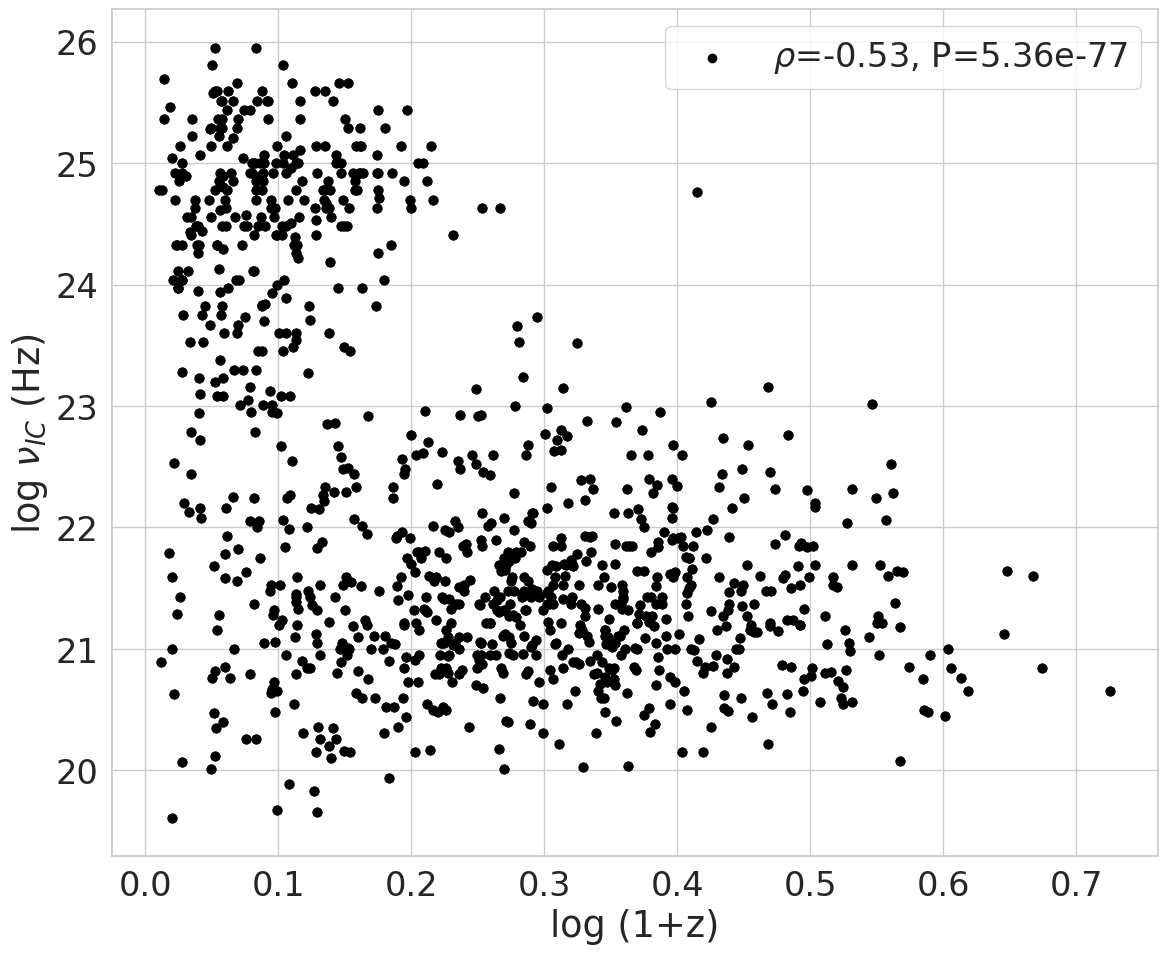}
	\includegraphics[width=0.4\linewidth]{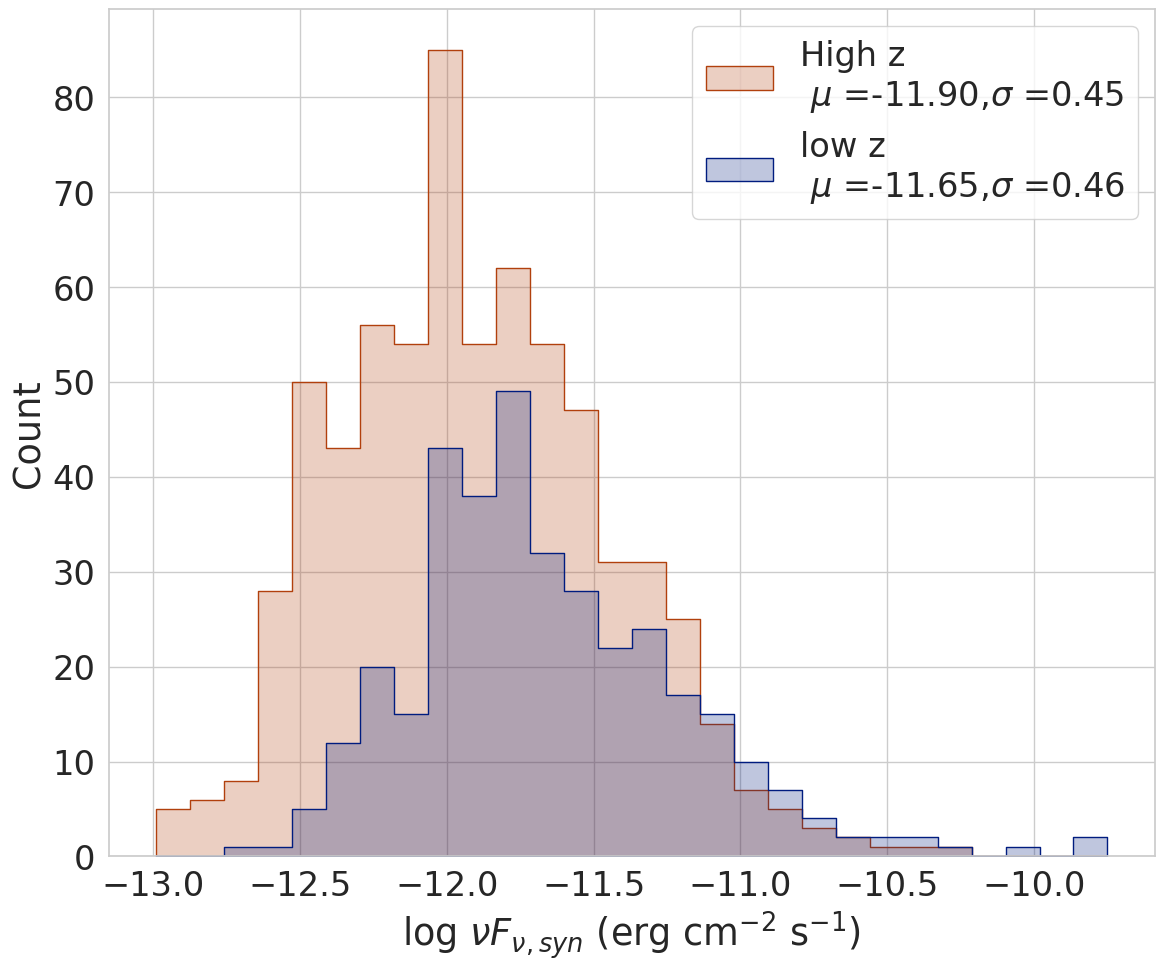}
	\includegraphics[width=0.4\linewidth]{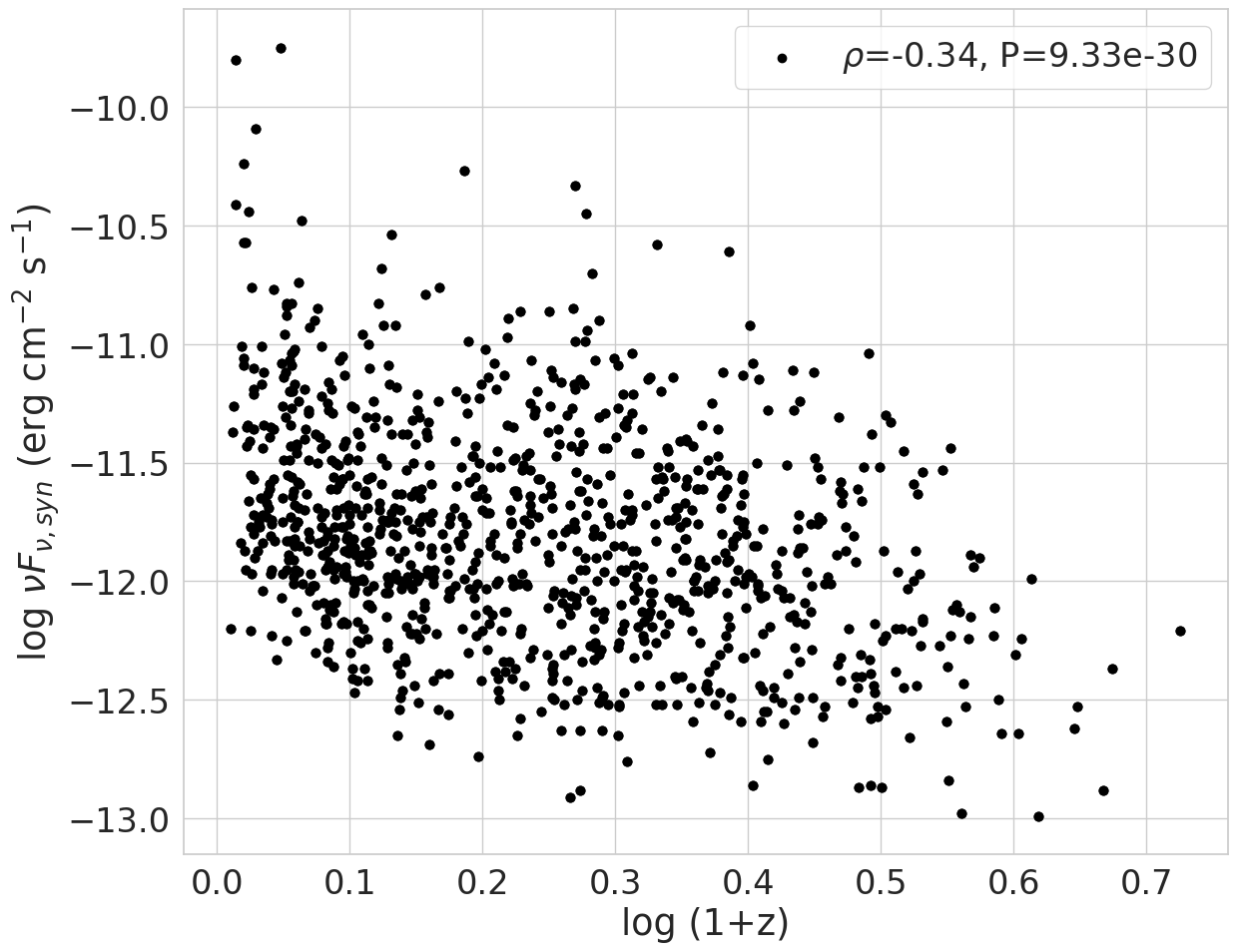}
	\includegraphics[width=0.4\linewidth]{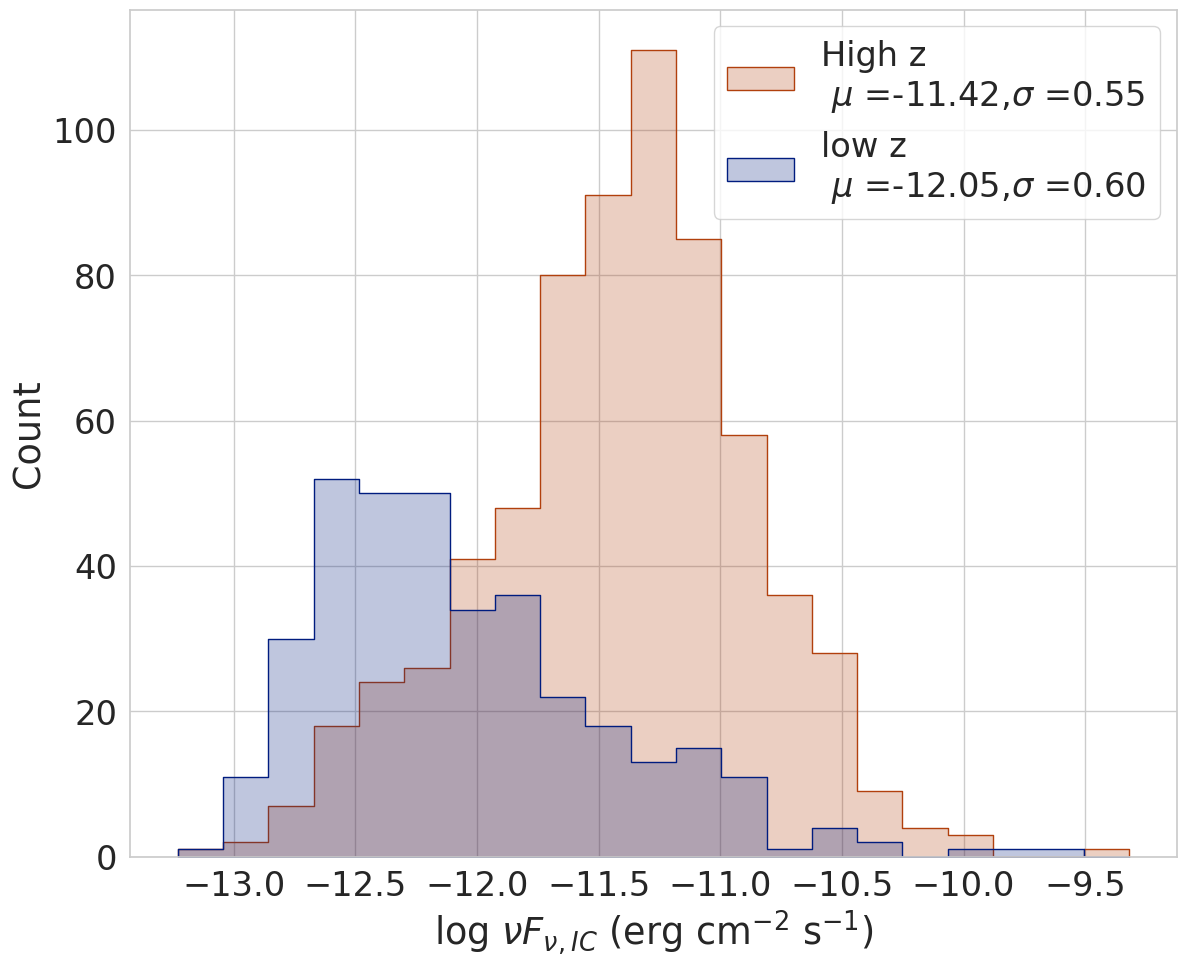}
	\includegraphics[width=0.4\linewidth]{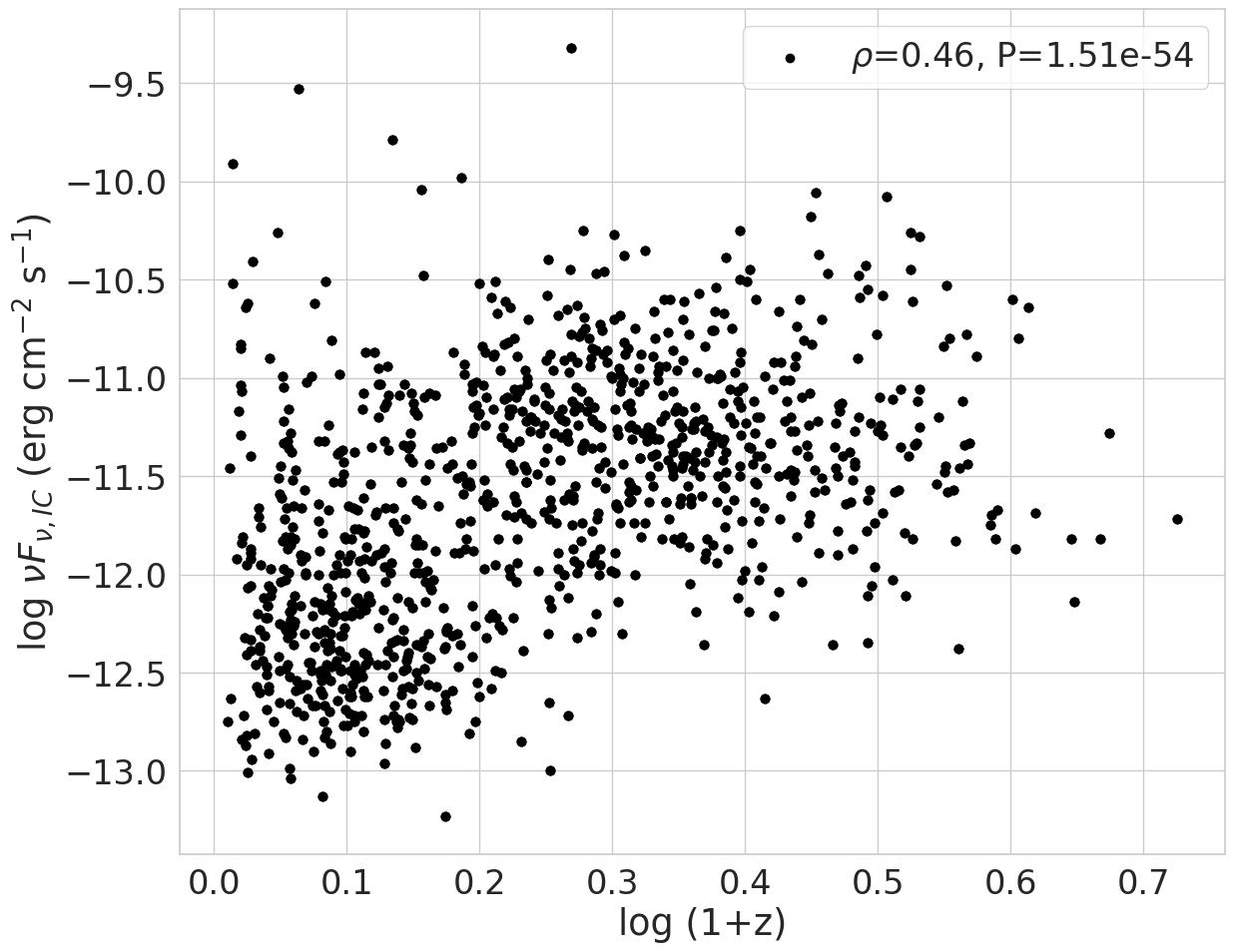}
	\includegraphics[width=0.4\linewidth]{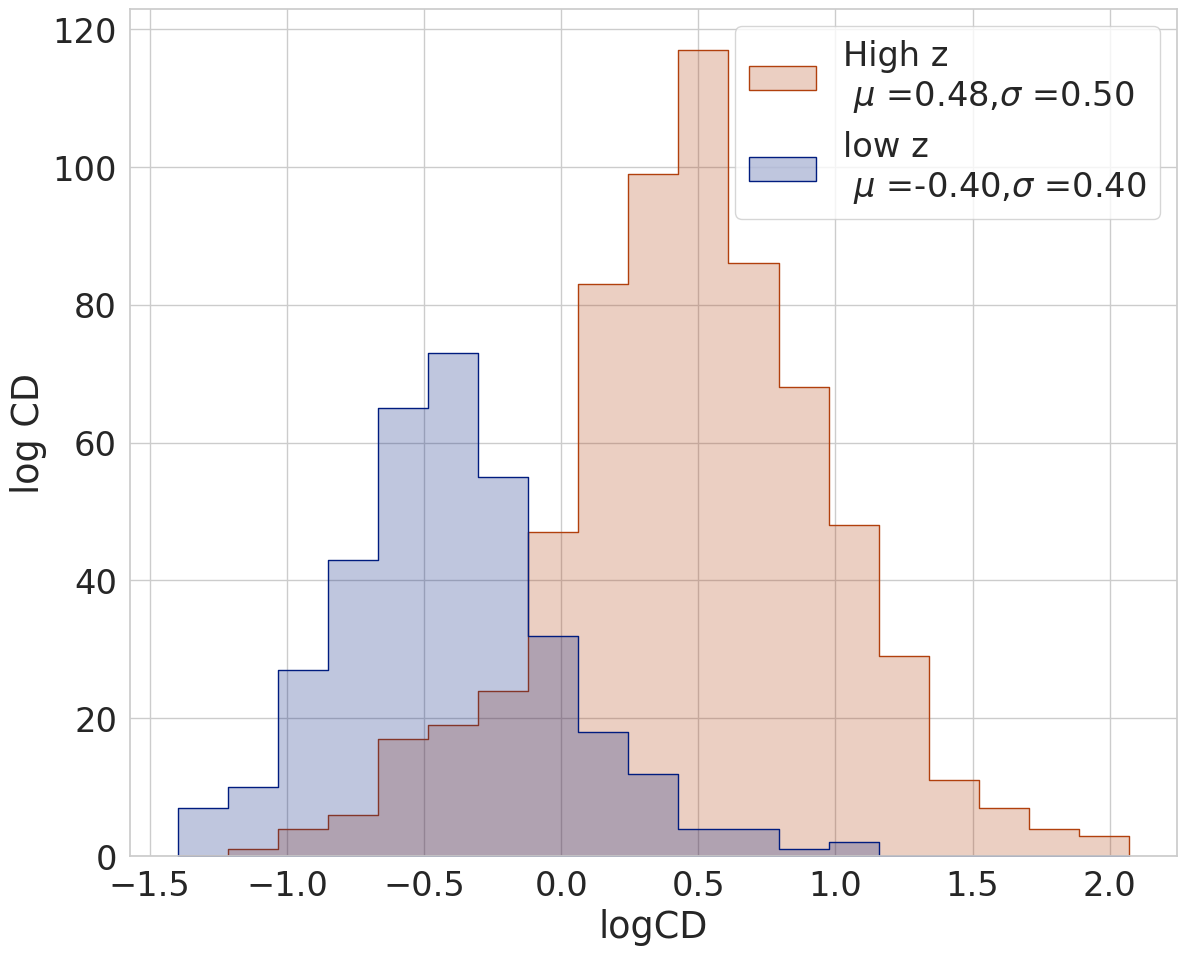}
	\hspace{2.5cm}\includegraphics[width=0.4\linewidth]{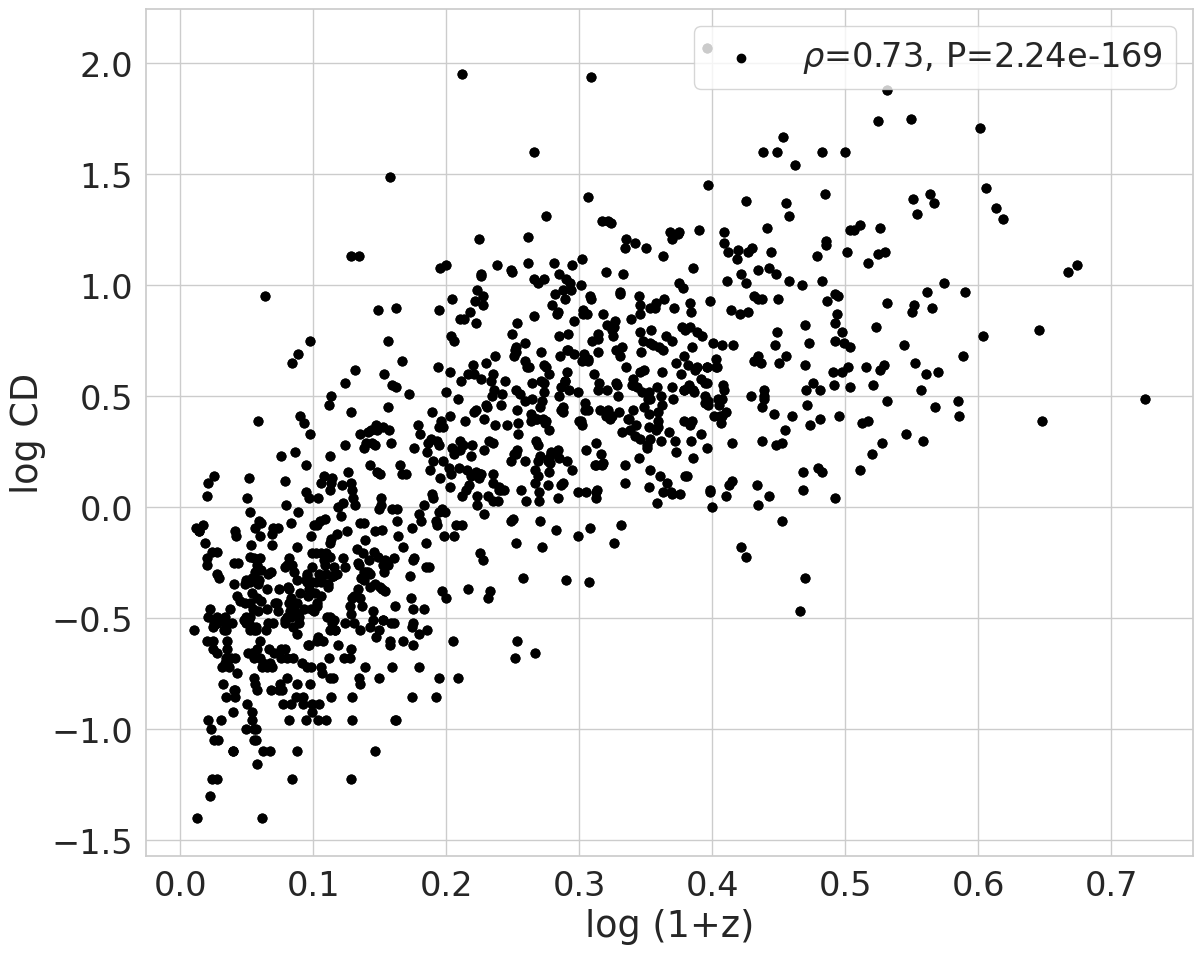}
	\caption{(Left) Histograms 
 \textls[-15]{of synchrotron peak frequency ($\nu_{syn}$), IC peak frequency ($\nu_{IC}$), energy fluxes at
	synchrotron ($\nu F_{\nu,syn}$) and IC ($\nu F_{\nu,IC}$) peak frequencies, and Compton dominance parameter ($CD$). 
	(Right)  Variations of log $\nu_{syn}$, log $\nu_{IC}$, log $\nu F_{\nu,syn}$, log $\nu F_{\nu,IC}$ and log $CD$ as a function of log (1 + z).}}
\label{fig:SED}
\end{figure}

\textls[-15]{We investigate the broadband SED properties using the sources in blazars central engines catalog. This catalog contains 
typical SED characteristics like  synchrotron peak frequency ($\nu_{syn}$), inverse Compton (IC) peak frequency ($\nu_{IC}$), 
energy fluxes at synchrotron ($\nu F_{\nu,syn}$) and IC ($\nu F_{\nu,IC}$) peak frequencies and Compton-Dominance parameter 
($CD= \nu F_{\nu,IC} / \nu F_{\nu,syn} $). For fair comparisons, we estimated $\nu_{syn}$ and $\nu_{IC}$ in the source frame by multiplying them with 
a factor (1 + z). The left panels in Figure~\ref{fig:SED} show the histogram distributions of $\nu_{syn}$, $\nu_{IC}$, $\nu F_{\nu,syn}$, $\nu F_{\nu,IC}$ 
and $CD$ for high and low-z blazars. The right panels in Figure~\ref{fig:SED} depict variations of these parameters with redshift (log (1 + z)). The 
two sample K-S test between high and low-z blazars results in \emph{p}-values of $\approx$0 for all the parameters. This suggests that the 
histogram distributions of these parameters are intrinsically different for high and low-z blazars.} \textls[-15]{A Gaussian function fit to the distributions of
$\nu_{syn}$, $\nu_{IC}$, $\nu F_{\nu,syn}$, $\nu F_{\nu,IC}$ and $CD$ gives $\mu \pm \sigma$ values of 13.42 $\pm$ 1.04 (15.32 $\pm$ 1.46), 
22.00 $\pm$ 1.07 (23.53 $\pm$ 1.68), $-$11.90 $\pm$ 0.45 ($-$11.65 $\pm$ 0.46), $-$11.42 $\pm$ 0.55 ($-$12.05 $\pm$ 0.60) and 0.48 $\pm$ 0.50 ($-$0.4 $\pm$ 0.40) for 
high-z (low-z) blazars, respectively. The Pearson correlation analysis indicates that $\nu_{syn}$ and $\nu_{IC}$ are anti-correlated with 
log (1 + z). Whereas, $\nu F_{\nu,IC}$ is positively correlated with log (1 + z) with $\rho$=0.46. However, a weak anti-correlation is observed 
between $\nu F_{\nu,syn}$ and log (1 + z) with $\rho$ = $-$0.34. $CD$ has a strong positive correlation with log (1 + z). This is broadly consistent with 
the \emph{blazar sequence} since at high-z only the higher luminosity objects can be detected, and their peak frequencies are shifted to lower 
values with increasing $CD$. Therefore, high-z blazars are mostly LSP blazars, and their HE hump typically peaks at MeV energies just below the 
energy range covered by the \emph{Fermi}-LAT. Their X-ray spectrum is harder and it usually corresponds to the rising part of the HE hump. 
Being strong and bright X-ray emitters, they are sometimes called \emph{MeV blazars}~\cite{Ghisellini2010a, Ghisellini2010}. The observations 
in the hard-X-ray band are ideal to detect the high-z blazars. Also, due to the low synchrotron peak frequency (typically in radio/IR waveband), the 
accretion disk emission is visible in the optical-UV region for most of the high-z blazars. Therefore, disk luminosity and black hole mass can be 
estimated by modeling the accretion disk emission~\cite{Shakura1973}. It is interesting to notice the variation of $\nu_{syn}$ with respect to 
log (1 + z) in the top right panel of Figure~\ref{fig:SED}. The low-z blazars show large variation in $\nu_{syn}$ (from 10$^{12}$ Hz to 10$^{19}$ Hz) 
making a vertical branch, whereas high-z blazars show little variation in $\nu_{syn}$ (from 10$^{12}$ to 10$^{14}$ Hz) making a horizontal branch in 
the scatter plot. Consequently, we have a bimodality in the top left panel of Figure~\ref{fig:SED}. This is in agreement with a recent blazar 
sequence study~\cite{Prandini2022} in which it has been observed that the synchrotron peak variation is mostly observed in BL Lacs and is insignificant 
in FSRQs. Since we mostly have BL Lacs at low-z and FSRQs at high-z in the data sample, a bimodality is obtained in the distributions of $\nu_{syn}$
and two branch feature in the variation of $\nu_{syn}$ with log (1 + z) in Figure~\ref{fig:SED}.}

%-------------------------------------Figure 4-------------------------------------------------

%--------------------------------------------------------------------

%-------------------------------------------Figure-5----------------------------------------------

%-------------------------------------------------------------------------
\subsection{Central Engine Properties}\label{sec:central_engines}
\begin{figure}[H]
	\includegraphics[width=0.49\linewidth]{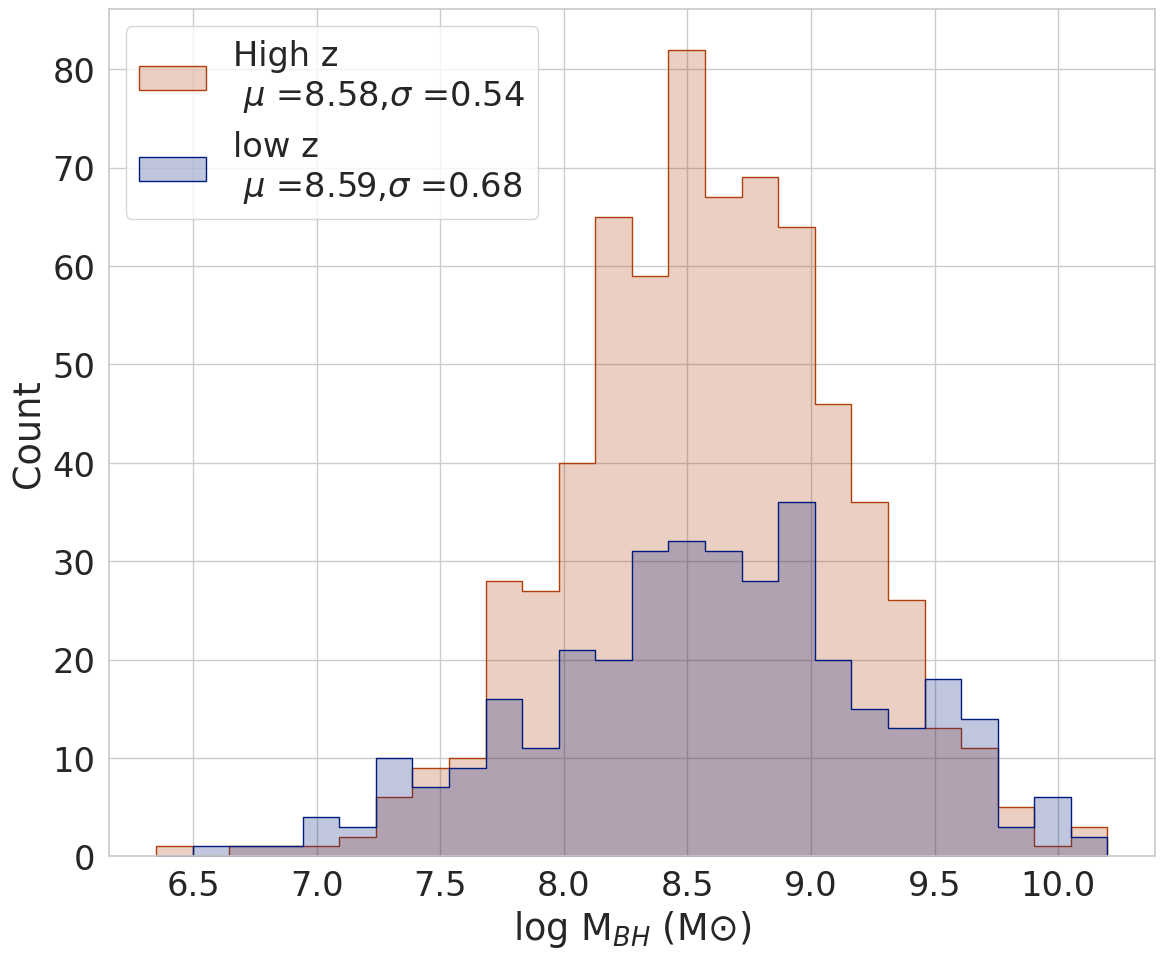}
	\includegraphics[width=0.49\linewidth]{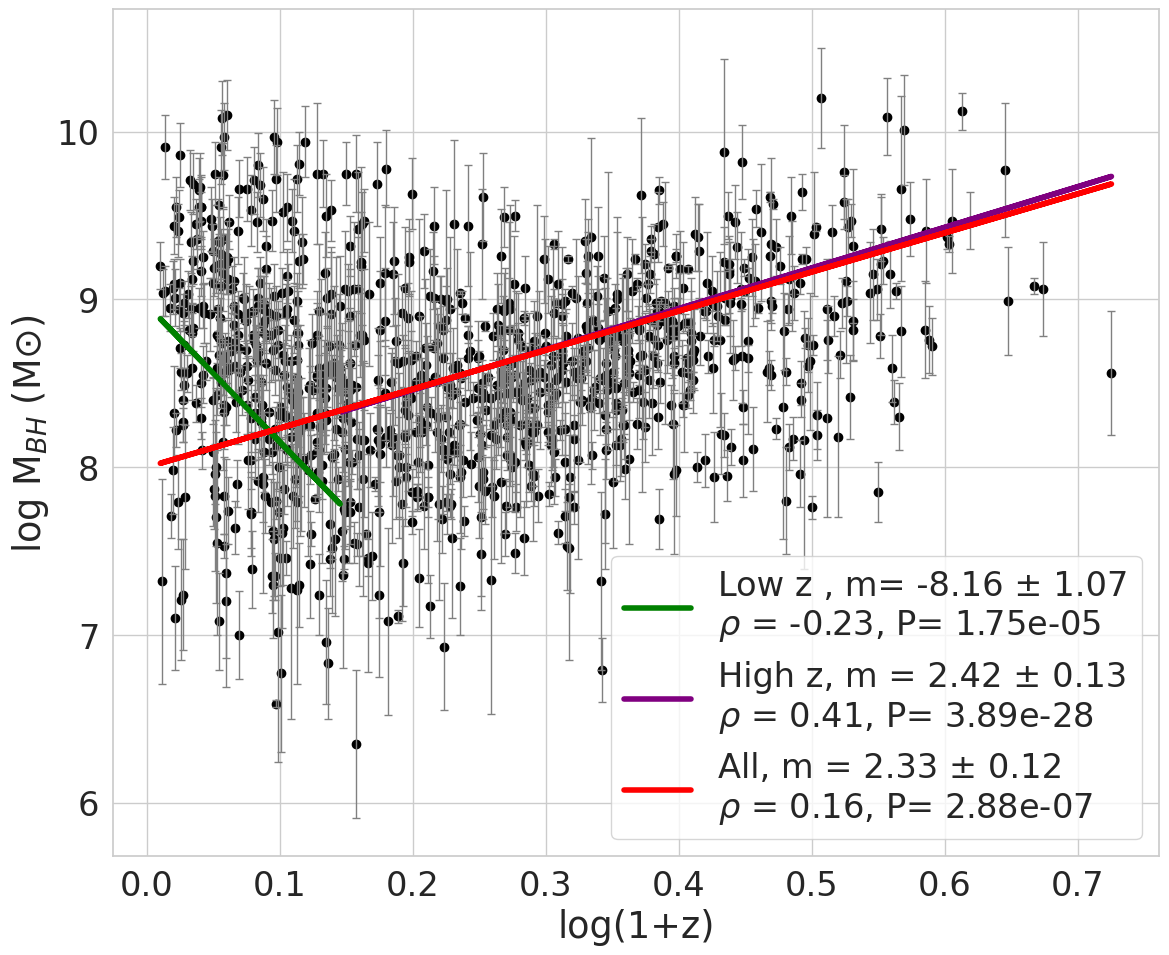}
	\caption{(Left) Histograms of log M$_{BH}$ values for high and low-z blazars. 
		 (Right) Scatterplot of log M$_{BH}$ and log (1 + z).}
	\label{fig:MBH}
\end{figure}

The blazar central engine consists of an SMBH at the center and an accretion disk surrounding it.  
It is characterized by the mass of SMBH (M$_{BH}$), accretion efficiency ($\eta$), and accretion disk luminosity ($L_{Disk}$). 
The sources in the blazar central engines catalog are considered for this study. The left panel in Figure~\ref{fig:MBH}
shows distribution of M$_{BH}$ values for high and low-z blazars. In both the cases, M$_{BH}$ values range from 
$\sim$2.5 $\times$ 10$^{6}$ M$_\odot$ to $\sim$1.75 $\times$ 10$^{10}$ M$_\odot$. The average M$_{BH}$ values for high and low-z 
blazars are similar with $\langle$log M$_{BH,\text{Low-z}} \rangle \approx \langle$log M$_{BH,\text{High-z}} \rangle \approx 8.6 $ 
and standard deviations are 0.54 and 0.68, respectively, when fitted with a Gaussian function. A two sample Kolmogorov--Smirnov (K-S) 
test is performed to see whether the two populations of M$_{BH}$ values are derived from the same parent population. The obtained \emph{p}-value 
of 0.11 suggests that there is a high probability both samples belong to the same parent source population. However, their correlations 
with $z$ are different as discussed below. 

The right panel in Figure~\ref{fig:MBH} shows the scatter plot of log $M_{BH}$ and log (1 + z) for high and low-z blazar populations. 
Overall, for the total blazar population, there is a very weak correlation between log M$_{BH}$ and log (1 + z) as indicated by $\rho =$ 0.16 
with \emph{p}-value of 2.88~$\times 10^{-7}$. However, by considering only high-z blazars, the correlation becomes relatively strong with 
$\rho=$ 0.41 and \emph{p}-value of 3.89$\times$10$^{-28}$. It suggests that more massive central black holes are associated with high-z blazars. 
This is likely due to selection bias in observations, as at high redshift only the most luminous (blazars with more massive black holes) sources 
can be detected. However, for low-z blazars, there is very weak anti-correlation between  log M$_{BH}$ and log (1 + z)  with $\rho$ = $-$0.23 and \emph{p}-value 
of  1.75 $\times$ 10 $^{-5}$. Although the K-S test suggests that there is a high probability that M$_{BH}$ values of low and high-z blazars 
belong to the same parent population, their correlations with z are different in low and high-z regimes. Therefore, M$_{BH}$ values of low and high-z 
blazars may not belong to the same \mbox{source population. }
\par
\begin{figure}[H]
	\includegraphics[width=0.49\linewidth]{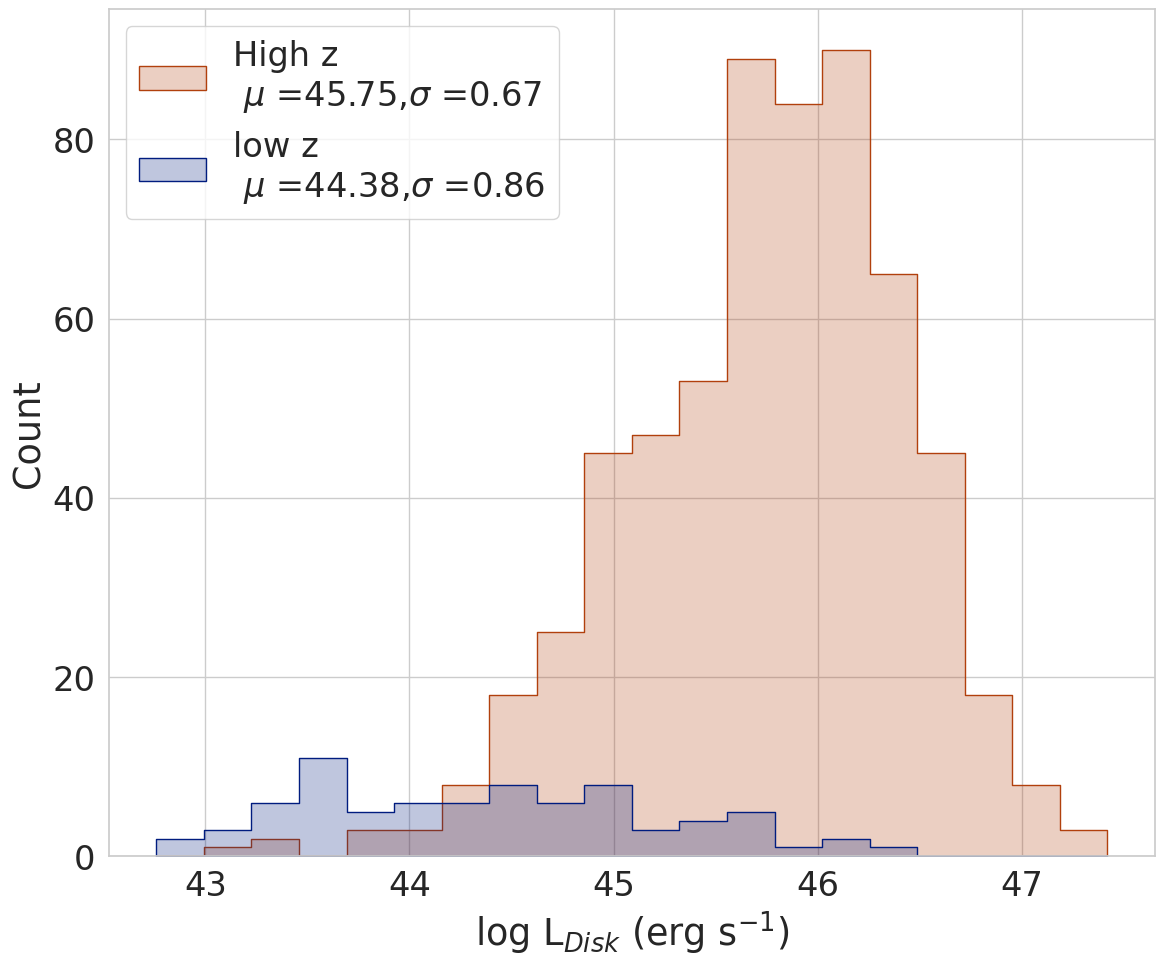}
	\includegraphics[width=0.49\linewidth]{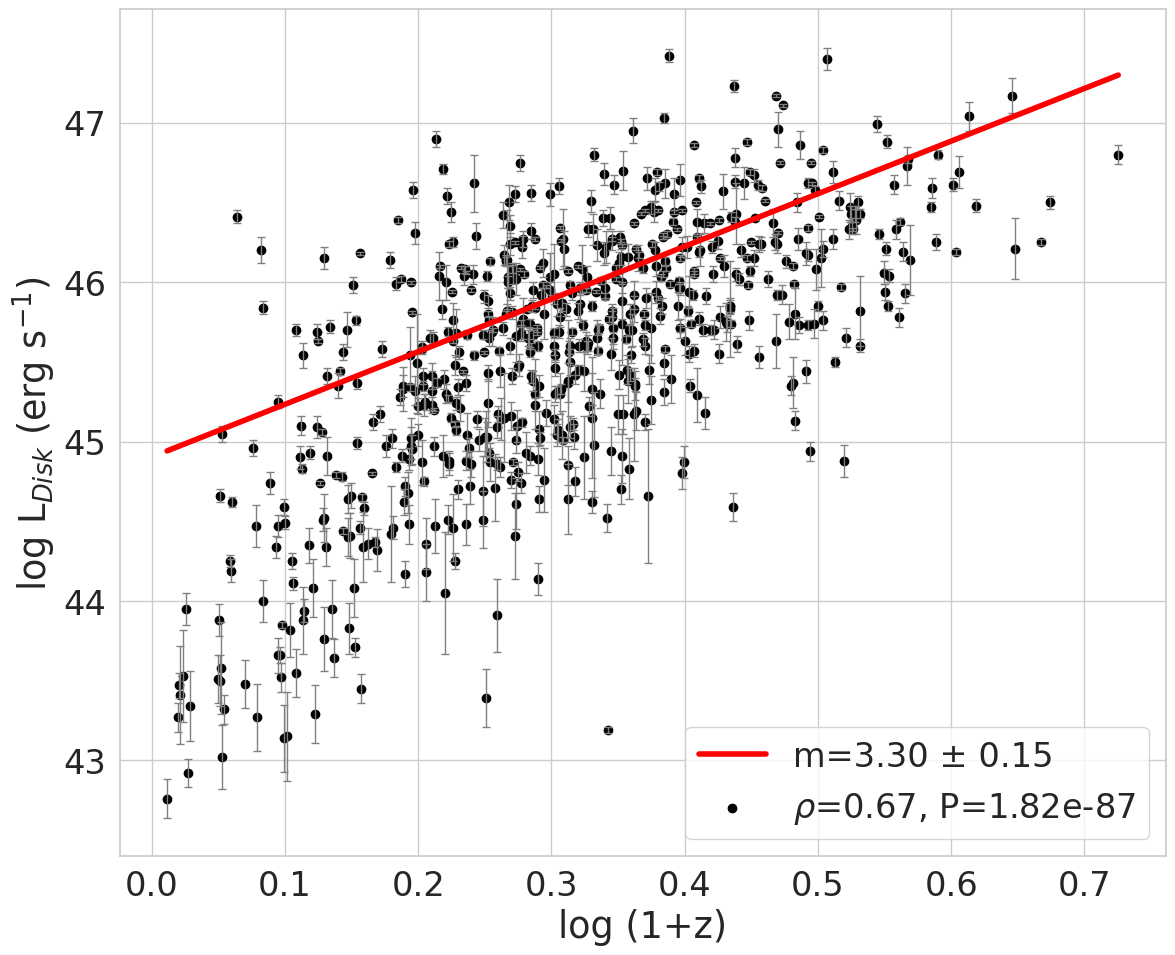}
	\includegraphics[width=0.49\linewidth]{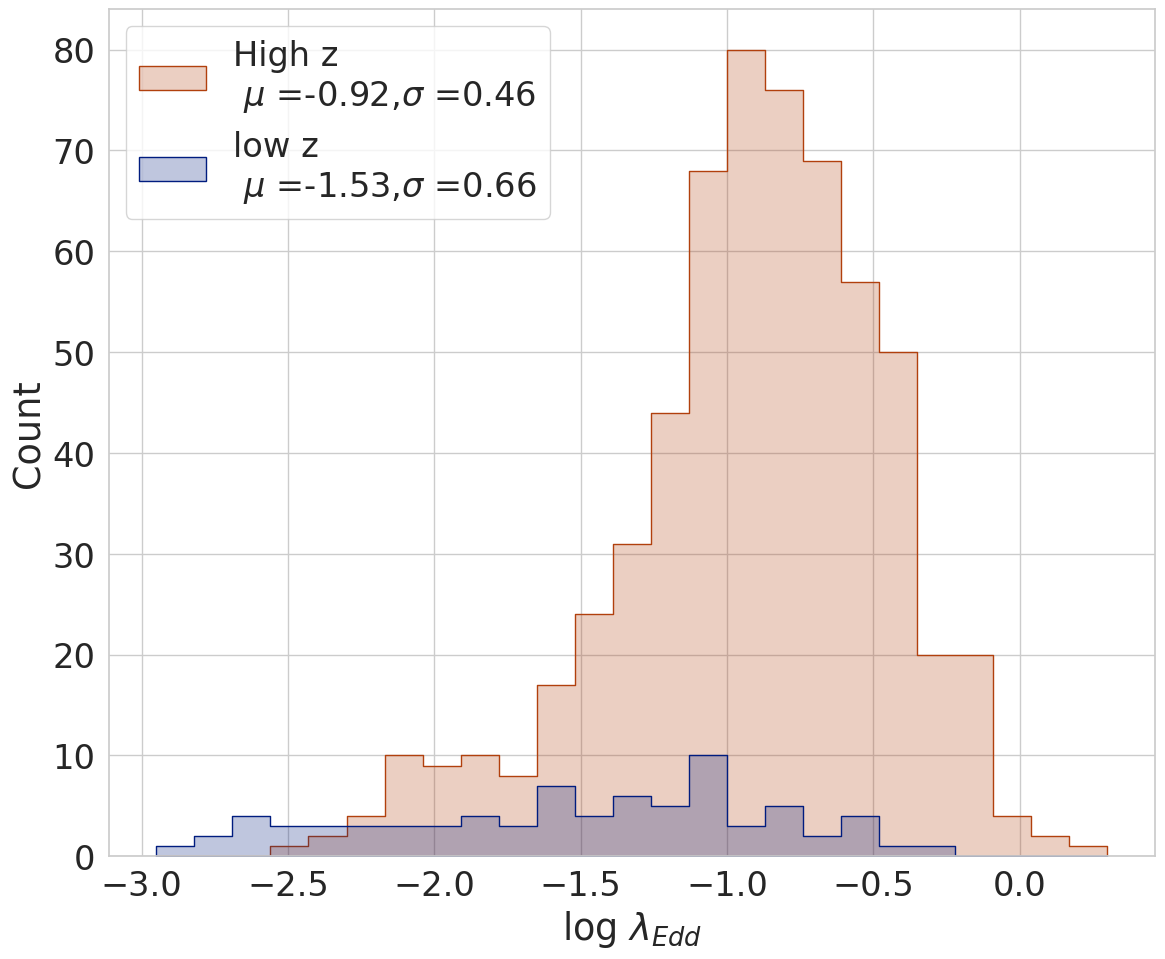}
	\includegraphics[width=0.49\linewidth]{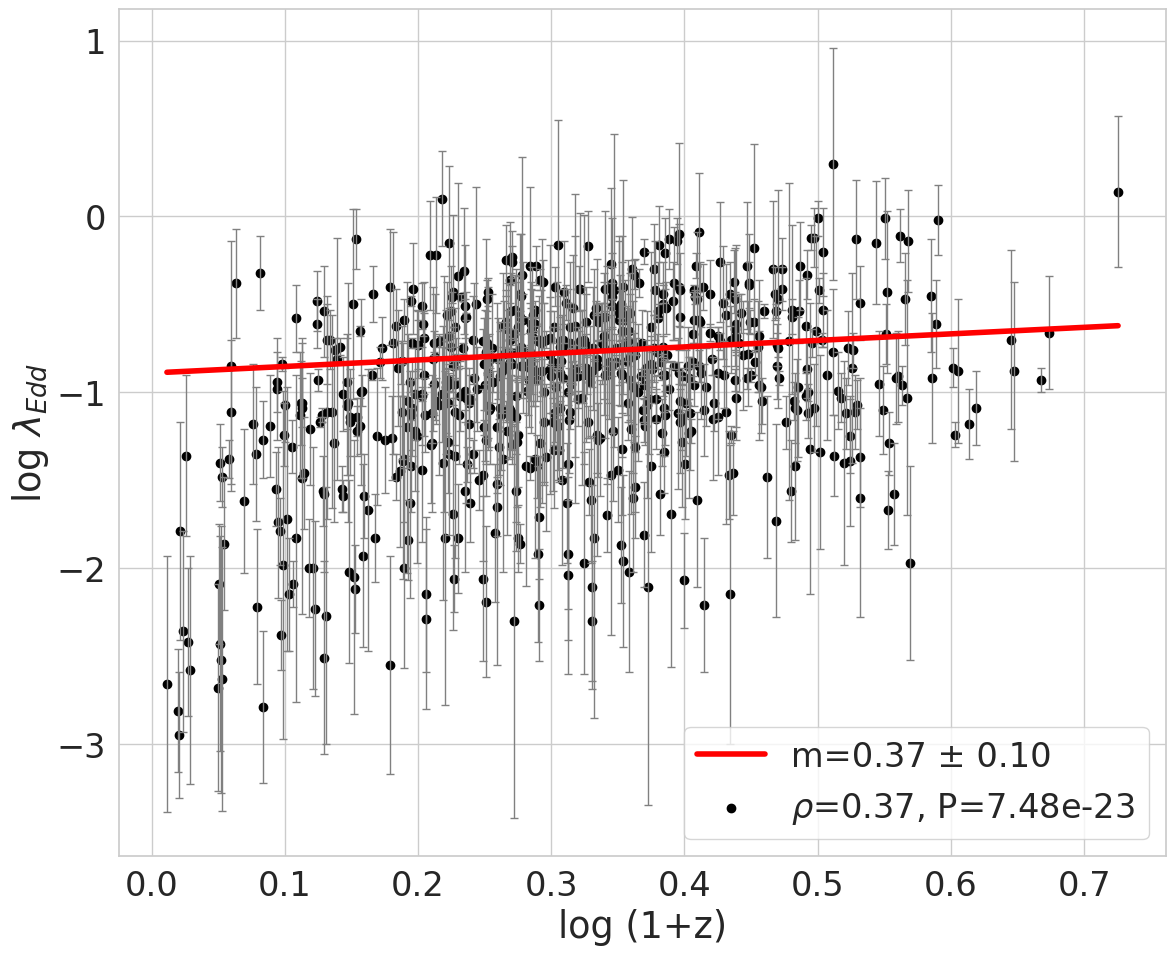}
	\caption{(Upper Left) Histograms of L$_{Disk}$ for high and low redshift blazars. 
	          (Upper Right) Scatter plot of log L$_{Disk}$ vs log (1 + z).
		  (Lower Left) Histograms of the Eddington ratios ($\lambda_{Edd}$) for high and low redshift blazars. 
		  (Lower Right) Scatter plot of log $\lambda_{Edd}$ as function of log (1 + z).}
	\label{fig:LDisk}
\end{figure}

\textls[-15]{To study the accretion disk properties, we consider the sources in the blazar central engines catalog. We have excluded 342 sources from this catalog  
as only upper limit values for $L_{Disk}$ are reported. The upper left panel in Figure~\ref{fig:LDisk} shows the histogram distributions of  
L$_{Disk}$ for blazars included in this study. \emph{p}-value of 2.08 $\times10^{-26}$  obtained from the two sample K-S test indicates that the distributions for 
low and high-z blazars are significantly different and belong to different parent distributions. A Gaussian function fitting to the distributions gives the 
mean values of log L$_{Disk}$ ($\langle$ log L$_{Disk} \rangle$) as 45.75 and 44.48 with standard deviations of 0.67 and 0.86 for high and low-z blazars, 
respectively. Therefore, in general, \mbox{high-z} blazars ($\langle$L$_{Disk,\text{High-z}} \rangle \approx 5.6 \times 10^{45} \text{erg} s^{-1}$)
have one order of magnitude higher accretion disk luminosity than low-z blazars ($\langle$L$_{Disk,\text{Low z}} \rangle \approx 3.0 \times 10^{44} 
\text{erg} s^{-1}$). The upper right panel in Figure~\ref{fig:LDisk} shows the variation of log L$_{disk}$ as a function of log (1 + z). The Pearson 
correlation analysis suggests a strong positive correlation between the two parameters with $\rho =$ 0.67 and \emph{p}-value of 1.28 $\times 10^{-82}$. 
Also, the linear regression analysis of the scatter plot gives the slope of line $m$ = 3.3 $\pm$ 0.15. Thus, the observed L$_{Disk}$ broadly 
varies as $\sim$ (1 + z)$^{ 3.3}$. }

%---------------------------------------Figure-6---------------------------------------------------

We also calculate the Eddington ratio ($\lambda_{Edd} = L_{Disk}/L_{Edd}$) using the definition of Eddington luminosity as 
L$_{Edd} = 1.26 \times 10^{38} (M_{BH}/M_{\odot}) \text{ erg } s^{-1} $. The corresponding distributions for high and low-z blazars 
and their variations as a function of redshift are shown in the lower left and lower right panels of Figure~\ref{fig:LDisk}, respectively. 
The $\lambda_{Edd}$ distributions for high and low-z blazars are intrinsically different as indicated by the K-S test \emph{p}-value 
of 8.74 $\times 10^{-13}$. The Gaussian function fitting to the distributions with mean and standard deviation ($\mu,~\sigma$) gives ($-$0.92,0.46) 
and ($-$1.53,0.66) for high and low-z blazars, respectively. This implies that high-z blazars 
accrete the surrounding matter with a high Eddington ratio of $\langle$ $\lambda_{Edd,\text{High-z}} \rangle = 0.12$ compared to 
the low-z blazars with $\langle$ $\lambda_{Edd,\text{Low-z}} \rangle = 0.03$. Therefore, high-z blazars have radiatively more efficient 
accretion disks than low-z blazars. This is also confirmed by the mild positive correlation obtained between $\lambda_{Edd}$ and log (1 + z)
with  $\rho$ = 0.37 and \emph{p}-value of $2.46 \times 10^{-4}$. The linear regression line in the scatter plot with m$=0.37 \pm 0.10$ 
suggests that observed $\lambda_{Edd}$ varies as $\sim$(1 + z)$^{\approx 0.37}$.
\par
Observation of blazars with strong luminosity, radiatively efficient accretion disk, and located at high redshift is likely connected 
with the cosmic evolution of blazars~\cite{Bottcher2002,Ghisellini2011,Ajello2014}. As discussed in Section~\ref{sec:z}, BL Lacs dominate 
the blazar population at low redshifts and FSRQs at high redshifts. According to the hypothesized evolutionary scenario, the FSRQs 
(likely high-z blazars) evolve into BL Lacs (likely low-z blazar) over cosmological time scales. \textls[-15]{At high-z, there is likely 
a gaseous matter-rich environment surrounding the central black hole for accretion, hence the higher values of $\lambda_{Edd}$. As they evolve 
into low-z blazars, the environment gradually gets depleted by accretion onto a black hole, the accretion disk becomes radiantly inefficient and 
accretion proceeds via advection-dominated flow~\cite{Narayan1997}. Alternatively, the observed features can also be attributed to the selection 
bias effects in observations as it is difficult to detect blazars having low $\lambda_{Edd}$ and low L$_{Disk}$ at high redshifts.}
%----------------------------------------------------------------------------------------
\subsection {Dynamics of Growth of SMBHs}\label{sec:Evolution}
High-z blazars provide a unique opportunity to explore the evolution of SMBHs and the role of relativistic jets in them~\cite{Volonteri2010,Volonteri2015}. 
Following the detailed calculations given in \cite{Shapiro2005,Belladitta2022}, we estimate the mass of black hole seed (M$_{BH,seed}$) required 
to produce currently observed SMBHs {at a} given $z$, assuming that it evolves with constant $\lambda_{Edd}$ and constant accretion efficiency defined as 
$\eta = L_{Disk}/\dot{M}_{BH}c^2$ during the entire evolutionary process. Accounting for the loss of accretion mass energy in the form of outgoing 
disk luminosity, the rate of M$_{BH}$ growth is given by 
\begin{equation}
	\frac{dM_{BH}}{dt} = (1-\eta) \dot{M}_{BH} = \frac{(1-\eta)\lambda_{Edd}L_{Edd}} {\eta c^2 }
	= \frac{(1-\eta)\lambda_{Edd}} {\eta } \frac{M_{BH}}{t_s}
\end{equation}
where t$_s$ is the characteristic accretion timescale (also known as Salpeter time) independent of M$_{BH}$. It is given as 
\begin{equation}
	t_s= \frac{\sigma_T c}{4 \pi G m_p} \approx 0.45 \text{ Gyr}
\end{equation}
Therefore, required black hole seed mass (M$_{BH,seed}$) for the formation of SMBH of mass M$_{BH}$ is given by
\begin{equation}
	\label{eqn:SMBH_Evolution}
	M_{BH,seed} =  M_{BH} \times \exp{\Bigl(\frac{-t}{\tau}\Bigr)}
\end{equation}
where $t$ is the total time available during which the seed black hole accretes and grows, $\tau$ is the characteristic $e-folding$ timescale 
defined as     
\begin{equation}
	\tau =  \frac{0.45 \eta}{\lambda_{Edd} (1-\eta)} \text{ Gyr}
\end{equation} 
\textls[-15]{The accretion 
 efficiency parameter $\eta$, believed to depend on the black hole spin, can vary between 0.1--0.3 or 0.4 for slowly to rapidly spinning 
black holes~\cite{Thorne1974,Shapiro2005}. We consider the sources in blazar central engines catalog for whom the classes (i.e., FSRQs, BL Lacs, or BCUs) 
are defined in the 4LAC-DR3 blazar catalog. For such blazars (558 FSRQs,53 BL Lacs, and 50 BCUs), we have calculated M$_{BH,seed}$ using 
Equation~(\ref{eqn:SMBH_Evolution}) by evolving them back to $z=30$ (corresponding to the Universe with age less than the first 100 million years 
when the first stars and galaxies are assumed to have formed)~\cite{Bromm2004,Bromm2011}. In our calculations, we use currently observed $\lambda_{Edd}$ 
throughout their evolution to see if such a scenario explains the observed SMBHs. The calculations are performed for two constant accretion efficiencies 
of $\eta$ = 0.1 and $\eta$ = 0.3~\cite{Shapiro2005}. For calculating \textit{t} in Equation~(\ref{eqn:SMBH_Evolution}), we use the standard flat 
$\Lambda$CDM model of the Universe with parameters H$_0$ = 70 km s$^{-1}$ Mpc$^{-1}$, $\Omega_M$ = 0.30 and $\Omega_\Lambda$ = 0.70. The final results depend on 
the assumptions made regarding various parameters like redshift of the seed formation, $\eta$  and $\lambda_{Edd}$.}
\par
\begin{figure}[H]
\vspace{-6pt}
	\includegraphics[width=0.5\linewidth]{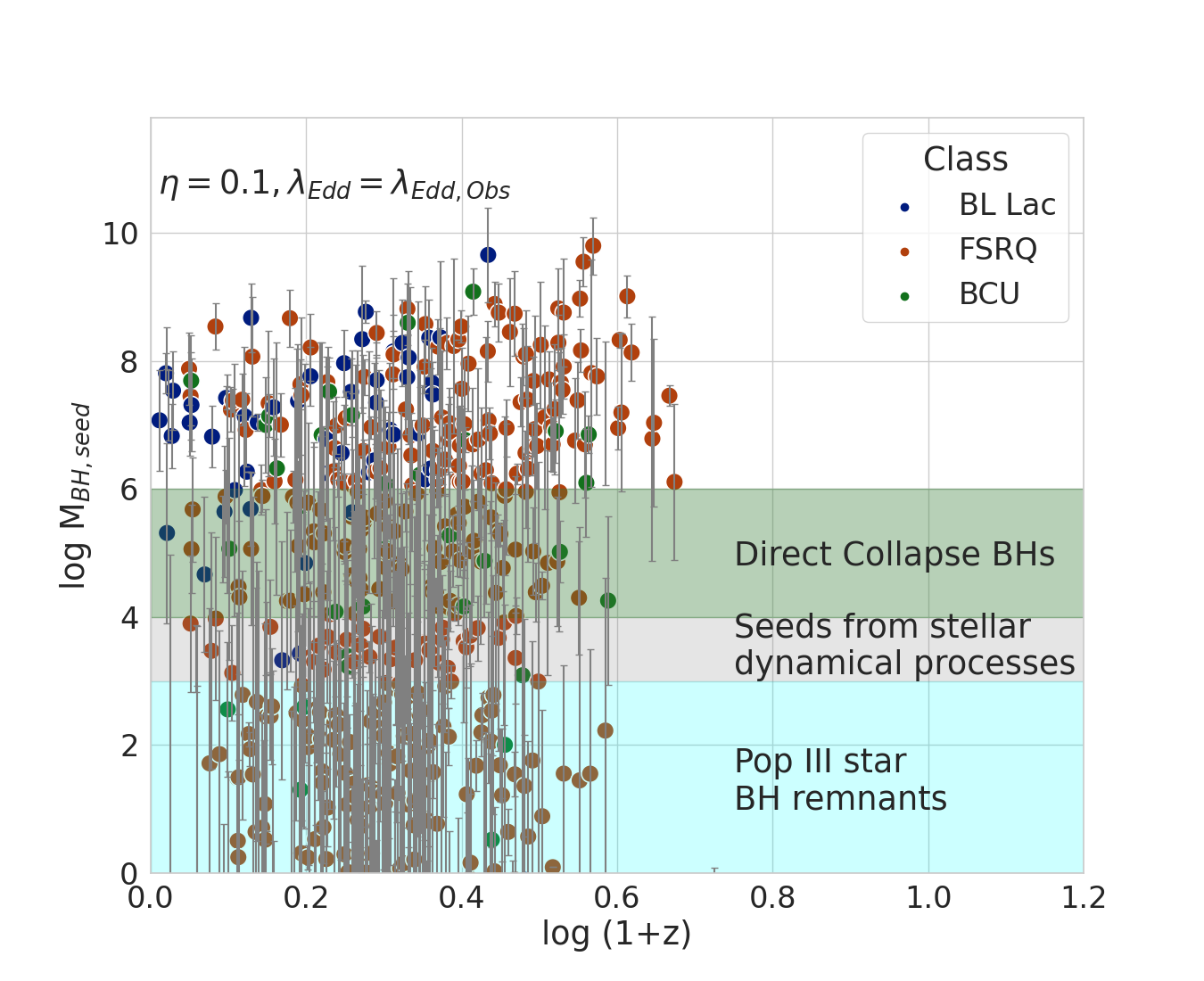}
	\includegraphics[width=0.5\linewidth]{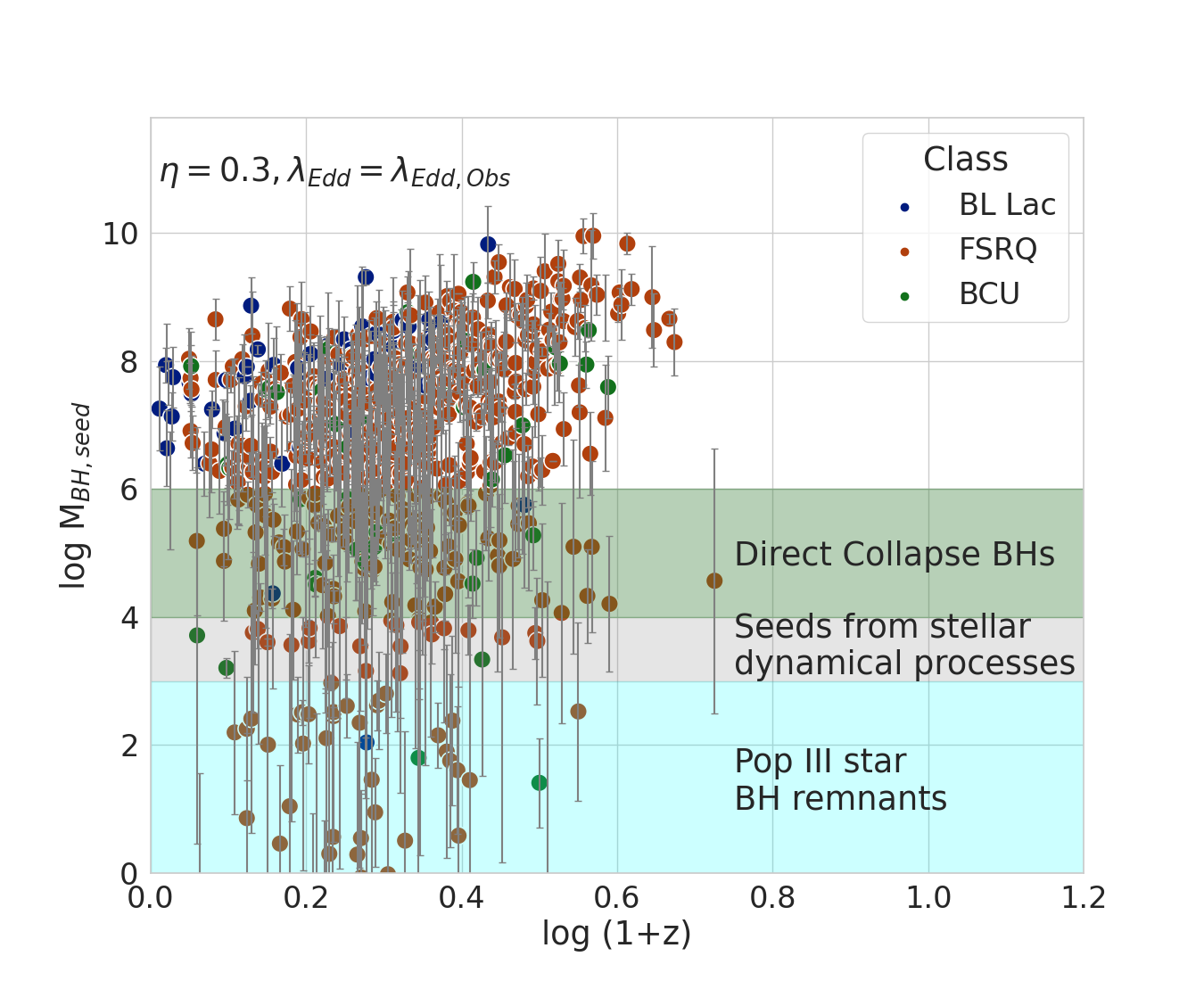}
	\caption{ \textls[-25]{Variations of the estimated seed M$_{BH,seed}$ values at z = 30, required for the formation of observed SMBHs in blazars, as a function 
	of log (1 + z), assuming the BH accretes with observed $\lambda_{Edd,obs}$ and with $\eta =$ 0.1 (left panel) and 0.3 (right panel).
	The shaded regions correspond to expected mass ranges of Population III star BH remnants (10--$10^3 \text{ M}_{\odot}$; cyan color),
	seed black holes from stellar dynamical processes ($ \ge 10^{3} \text{ M}_{\odot}$; gray color) and black holes from direct collapse 
	($10^4$--$10^6 \text{ M}_{\odot} $; green color).}}
	\label{fig:MBHSeed}
\end{figure}

\textls[-15]{Figure~\ref{fig:MBHSeed} shows the variations of the estimated values of M$_{BH,seed}$ as a function of log (1 + z) for $\eta = 0.1 $ (left panel) and 
$\eta = 0.3$ (right panel). In these plots, we also show the expected mass ranges of different astrophysical objects present at $z=30$, which could have been 
the seeds for the formation of observed SMBHs in the centers of galaxies. They are (i) population III star black hole remnants: 10--10$^3 M_{\odot}$ (cyan color) 
\cite{Abel2000,Bond1984}, (ii) seeds from stellar dynamical processes: $10^3-10^4 M_{\odot}$ (gray color)~\cite{Devecchi2009,Lupi2014,Alexander2014,Volonteri2016,Lupi2016}, 
and (iii) direct collapse black holes: $10^4$--$10^6 M_{\odot}$ (green color)~\cite{Haehnelt1993,Begelman2006}. The presence of SMBHs in blazars corresponding to 
$M_{BH,seed}$ values lying below the horizontal line $log M_{BH,seed} = 6$ in Figure~\ref{fig:MBHSeed} can be explained by the evolution of one of the 3 astrophysical 
seed objects mentioned above. However, a significant number of blazars lie above the line $log M_{BH,seed} = 6$, and plausible explanations for this are given below.}

%-----------------------------Figure-7---------------------------------------------------------------------

The left panel in Figure~\ref{fig:MBHSeed} indicates that, for $\eta = 0.1 $, about 26\% of the blazar population requires more massive progenitors than what direct 
collapse models or stellar dynamical processes predict. This fraction goes up to $\sim$69\% when we consider $\eta = 0.3$ (right panel of Figure~\ref{fig:MBHSeed}). 
Therefore, high values of accretion efficiencies, i.e., high conversion rate of rest mass energy to luminous energy, are not realistic to explain the observed SMBHs. 
This is consistent with the similar conclusions derived in the literature~\cite{Shapiro2005}. Alternatively, these blazars must accrete the matter with a higher Eddington 
ratio than the currently observed $\lambda_{Edd}$ for a significant time during their evolution to explain the growth of observed SMBHs. Thus, the higher the $\eta$, the higher the $\lambda_{Edd}$ we need to explain the growth of the SMBHs. A major fraction of the SMBHs in BL Lacs ($\sim$75\% and $\sim$92\% of the total BL Lacs 
corresponding to $\eta =$  0.1 and 0.3, respectively) require seed mass more than the expected seed mass from direct collapse models. This can be due to their 
observed low accretion disk luminosity and hence low Eddington ratio ($\lambda_{Edd}$). Therefore, these objects must have accreted the matter with 
higher $\lambda_{Edd}$ during their evolution to explain the observed SMBHs. This implies that there was a conducive gas-rich environment around the central engines 
in their earlier growth time (high-z). These objects would have radiatively efficient accretion disks in their adulthood, meaning they were in the FSRQ phase 
(high $\lambda_{Edd}$ phase) and as their gas-rich environment became depleted, their accretion disk became radiatively inefficient and they evolved into BL Lacs 
at relatively low-z. It is evident from the figure that a significant number of high-z blazars (or radio-loud AGN) require higher mass progenitors than the currently 
hypothesized seed populations. Therefore, we need higher $\lambda_{Edd}$ or sometimes the occurrence of super-Eddington episodes to explain the growth of SMBHs. Such 
scenarios are invoked to explain the growth of SMBHs in the recently discovered high redshift AGNs J1205-0000 (z = 6.699)~\cite{Onoue2019} and PSO J006+39 (z = 6.621) 
\cite{Tang2019}. Although there are possibilities for super Eddington accretion episodes in gas-rich or dusty strong wind environments near the 
central engines~\cite{Kim2015,Kubota2019}, whether such episodes are sustainable or not for a significant time is an open question in the study 
of the growth of SMBHs at high redshifts. Another possible scenario has been proposed in the literature, where in the presence of a jet, not all the gravitation 
potential energy of in-falling matter is converted into luminous energy but some part of it can go to increase the magnetic field energy inside the 
jet, which is necessary for jet launching processes~\cite{Blandford1977}. Therefore, for a given accretion efficiency, there will be fewer luminous disks and hence 
lower ($\lambda_{Edd}$) compared to the normal accretion scenario. This implies that, in the presence of a jet, black holes grow faster with a large accretion rate 
without increasing the disk luminosity ~\cite{Jolley2008,Jolley2009,Ghisellini2010a}. Cosmological simulations were also been explored to investigate the 
growth of seed black holes through processes other than accretion such as merger and AGN feedback~\cite{Alexander2012,Feng2014}. Similar calculations
and conclusions have also been reported in a recent study~\cite{Belladitta2022}. Hence, detection and detailed study 
of high-z blazars are important to understand the growth of SMBHs and their connection with relativistic jets. It is important to note that the conclusions 
drawn here rely on the reported values of $\Lambda_{Edd}$, L$_{Disk}$, and M$_{BH}$ in the blazar central engines catalog. These values were derived 
from scaling relations using single-epoch optical spectra. They are indirect methods and have inherent assumptions and limitations which may lead to 
significant bias~\cite{Shen2011}. Therefore, the estimation of M$_{BH}$ from different approaches like reverberation mapping, etc., are crucial 
\cite{Peterson1993,Singh2022}.
%------------------------------------------------------------------------------------------------------
\subsection{Properties of Emission Region}\label{sec:emission_zone}
\begin{figure}[H]
	\includegraphics[width=0.48\linewidth]{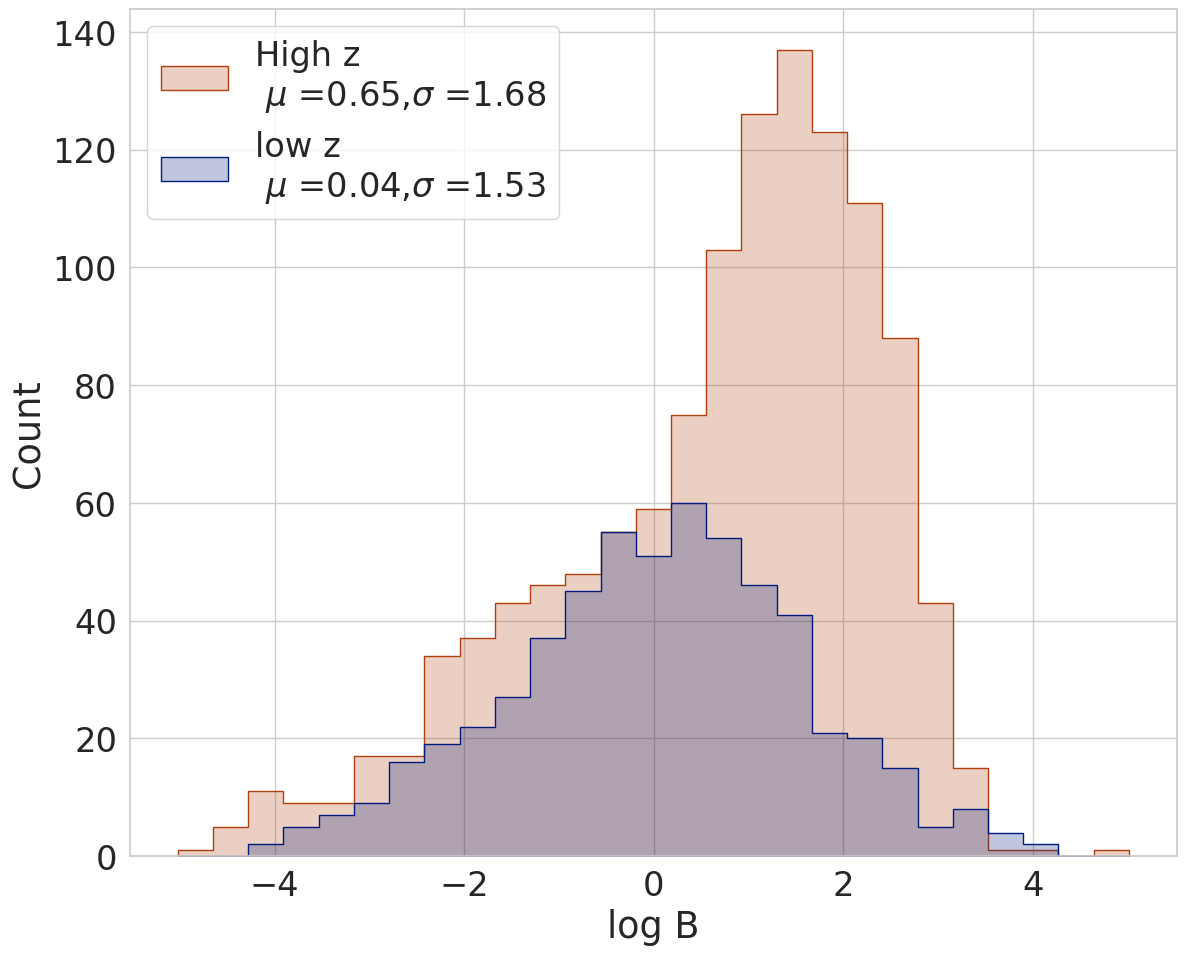}
	\includegraphics[width=0.48\linewidth]{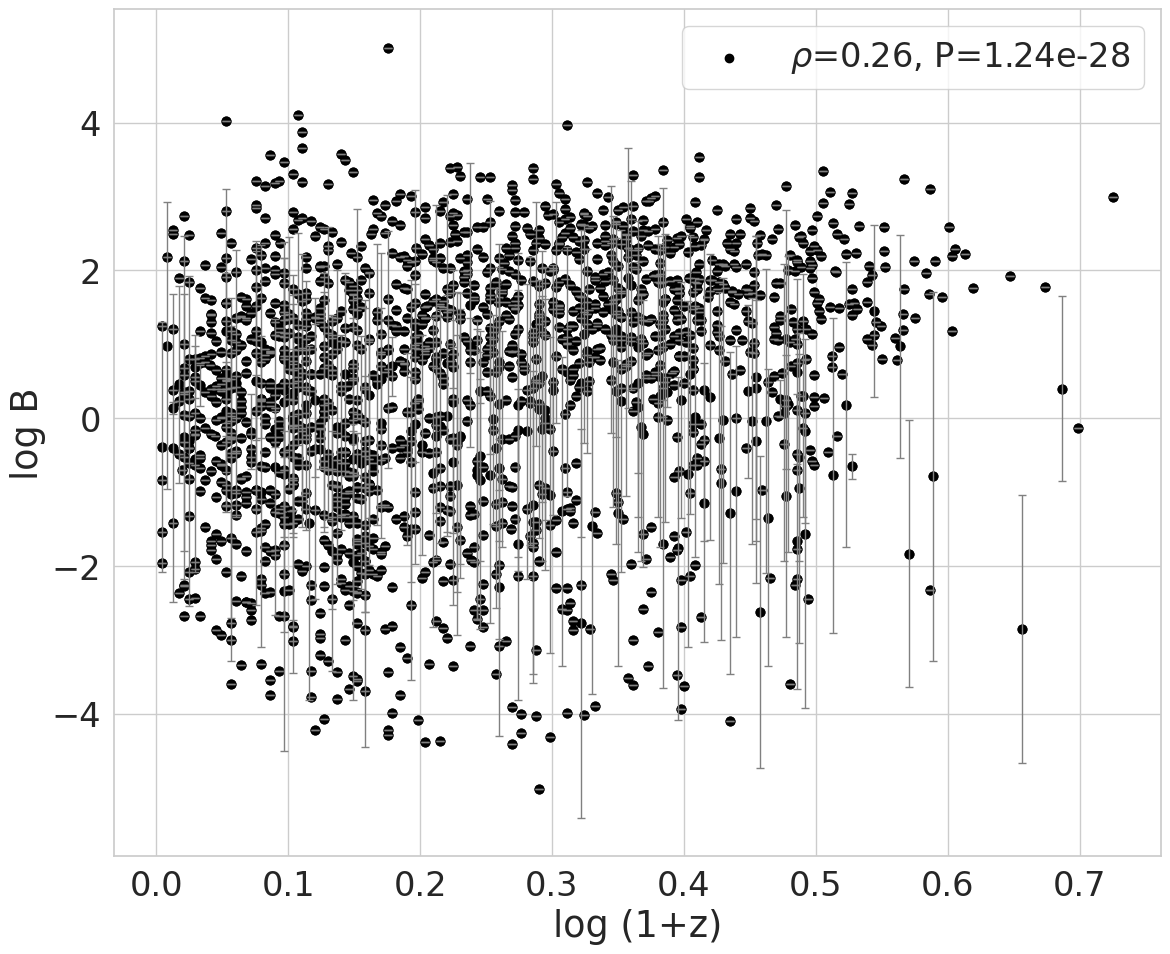}
	\includegraphics[width=0.49\linewidth]{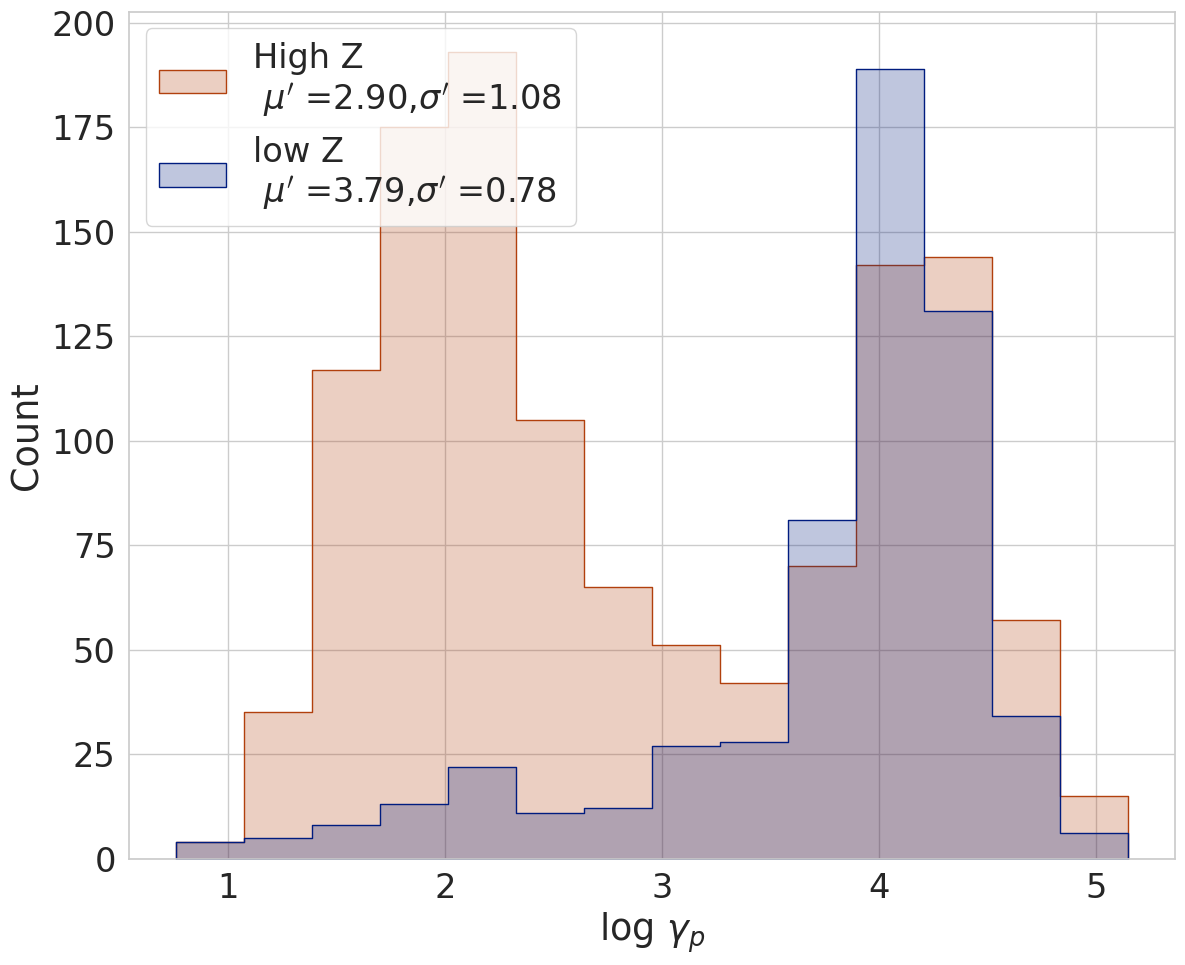}
	\includegraphics[width=0.49\linewidth]{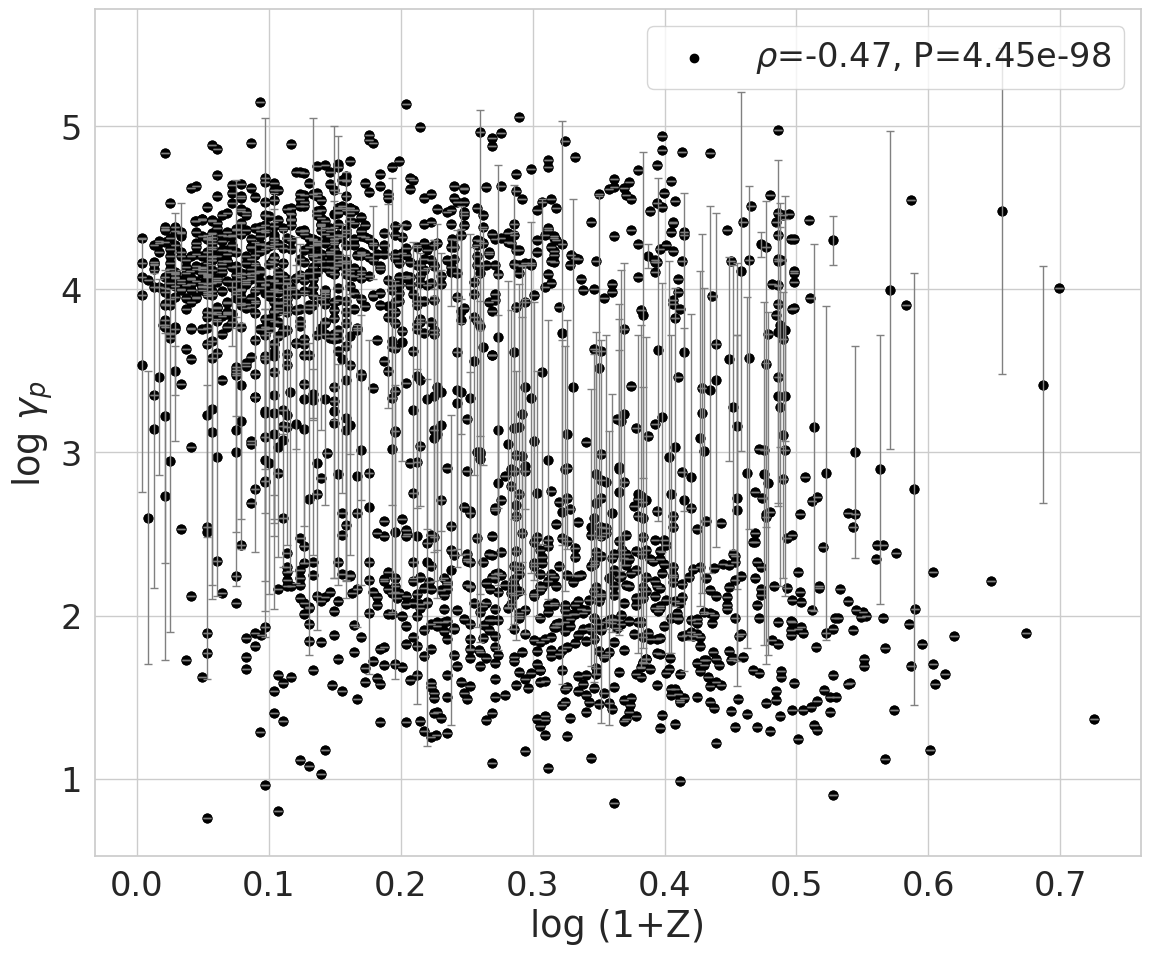}
	\caption{(Left) Histogramsof log B and log $\gamma_p$ for high and low redshift blazars. 
	          (Right) Variation of log B and log $\gamma_p$ values with log (1 + z).}
	\label{fig:emission_zone}
\end{figure}

\textls[-15]{In the one-zone leptonic emission model, the emission region is mainly characterized by its magnetic field $B$ and relativistic electron energy distribution having 
peak Lorentz factor $\gamma_p$. We use log B and log $\gamma_p$ parameters reported in the blazar emission regions catalog for this study. For some sources, 
ranges of log B and log $\gamma_p$ are reported in the catalog, and we take the midpoint of the range as the value for log B and log $\gamma_p$ for them.
The left panels in Figure~\ref{fig:emission_zone} show the histograms of log B and log $\gamma_p$ for high and low-z blazars. The right panels show the variations 
of these parameters as a function of log (1 + z). \emph{p}-values of two sample K-S tests for high and low-z distributions of log B and log $\gamma_p$ parameters are close 
to 0. This suggests that the distributions of B and $\gamma_p$ for high and low-z blazars are intrinsically different and belong to the same parent distribution. 
A Gaussian function fitting to the histograms of log B gives $\mu \pm \sigma$ of 0.65 $\pm$ 1.68 (B = 4.46 G) and 0.04 $\pm$ 1.53 (B = 1.09 G) for high and low-z 
blazars, respectively. Also, there exists a mild positive correlation between B values and log (1 + z) with $\rho$ = 0.26 and the P value of 1.24$\times 10 ^{-28}$. 
Therefore, high-z blazars require emission zones with higher magnetic fields compared to low-z blazars. The histogram distribution of log $\gamma_p$ for 
high-z blazars shows a clear double hump structure with peaks around log $\gamma_p$ values of 2 and 4. Similarly, the scatter plot of log $\gamma_p$ and log (1 + z) 
shows two horizontal branches around these peak values. A careful analysis suggests that these peak values actually correspond to the peak values of distributions 
of log $\gamma_p$ for all FSRQs and BL Lacs. Overall, two peaks in the histogram distributions and two branches in the scatter plots actually correspond to the FSRQs 
and BL Lacs, respectively. Instead of fitting the Gaussian profile to the histograms of log $\gamma_p$, we calculated the arithmetic mean ($\mu^\prime$) and standard 
deviation ($\sigma^\prime$) for the distributions. The $\mu^\prime$ values for high and low-z blazars are found to be 2.90 and 3.79 corresponding to peak Lorentz factors 
of 7.94 $\times 10^2$ and 6.17 $\times 10^3$, respectively. The scatter plot shows two distinct branches corresponding to the FSRQs and BL Lacs, and there is also a 
mild anti-correlation between log $\gamma_p$ and log (1 + z) with $\rho$ = $-$0.47. These results are consistent with the blazar emission models. At high redshifts, 
most of the sources are FSRQs and BL Lacs dominate the population at low redshifts. FSRQs contain photon-rich environments like BLR and DT owing to radiatively efficient 
accretion disk, and relativistic electrons quickly lose their energy by interacting with ambient photons via the IC process. Therefore, these blazars have high Compton 
dominance, and the electron population has low $\gamma_p$. Consequently, we observe blazars with low synchrotron frequency peaked SEDs. On the other hand, BL Lacs have a 
photon-starved environment surrounding them. Relativistic electrons can attain high $\gamma_p$ in the absence of target photon fields. Therefore, their SED peak 
lies at higher frequencies with very low Compton dominance ~\cite{Bottcher2002,Ghisellini2011,Paliya2021,Fan2023}. The variation of log $\gamma_p$ as a function of 
log (1 + z) also contributes to the observed behavior of spectral parameters measured by the \emph{Fermi}-LAT. The higher the z, the higher the efficiency of the disk, and the higher the 
density of external photon fields, resulting in the electron energy distributions with lower $\gamma_p$. Therefore, this can also contribute to the observed soft photon 
spectral indices and higher curvatures for high-z blazars.}
%-------------------------------------Figure-8----------------------------------------------

%-------------------------------------Section-5-Summary and Outlook---------------------------------------
\section{Summary}\label{summary}
In this work, we studied the properties of high-z (z $>$ 0.4) blazars using the data available in the literature. We compared the observed properties of SEDs 
and derived properties of the central engines and emission zones of high and low redshift blazars. A summary of the main results obtained in this study is 
given below:
\begin{itemize}
	\item	Redshift distribution study of 4 LAC-DR3 blazars catalog suggests that the FSRQs dominate the high-z blazar population with 
		62\% composition followed by BL Lacs and BCUs with 30\% and 8\% compositions, respectively. However, in the low-z regime, BL Lacs 
		dominate the total blazar population with 80\% composition and FSRQs (13\%) along with BCUs (7\%) being the minority. The reasons for 
		these anomalies are likely due to observational biases and/or due to hypothesized cosmic blazar evolution theories.

	\item   \textls[-25]{Observed \emph{Fermi}-LAT photon spectra for the high redshift blazars are found to be softer with higher curvature compared to 
		low redshift blazars. The histograms of PL index ($\Gamma$), LP slope at 1 GeV/(1 + z) ($\alpha^\prime$)  
		and LP curvature ($\beta$) roughly follow Gaussian distributions with $\mu \pm \sigma$ values of $\Gamma = 2.33\pm 0.25 (2.02 \pm 0.23)$,
		$\alpha^\prime = 2.10 \pm 0.39 (1.84 \pm 0.38)$ and $\beta = 0.11 \pm 0.07 (0.07 \pm 0.07)$ for high (low) redshift blazars. Also, $\Gamma$, 
		$\alpha^\prime$ and $\beta$ exhibit a positive correlation with log (1 + z). This is likely due to the fact that high-z blazars suffer more 
		EBL attenuation of HE and VHE $\gamma$ photons, leading to softer $\gamma$-ray spectra than low-z blazars. Also, different spectral shapes of  
		the underlying energy distribution of emitting particles in low and high-z blazars contribute to the observed \emph{Fermi}-LAT $\gamma$-ray 
		spectral features.}

	\item   Observationally, high-z blazars are the most luminous objects in the $\gamma$-ray Universe. The mean $\gamma-$ray, X-ray, optical and radio luminosities 
		(in $erg~s^{-1}$) of high (low) redshift blazars are estimated as $\langle L_\gamma \rangle = 6.31 \times 10^{45} (6.03 \times 10^{43})$, 
		$\langle L_X \rangle = 1.58 \times 10^{45} (1.02 \times 10^{44})$, $\langle L_O \rangle = 4.46 \times 10^{45} (2.57 \times 10^{44})$ and
		$\langle L_R \rangle = 8.13 \times 10^{42} (1.0 \times 10^{41})$, respectively. These luminosities have a strong positive correlation with the luminosity 
		distance (D$_L$) with functional dependencies of $L_\gamma \propto D_{L}^{2.27}$, $L_X \propto D_{L}^{1.47}$, $L_O \propto D_{L}^{1.51}$ and 
		$L_R \propto D_{L}^{2.14}$. 

	\item   The investigation of broadband SED properties reveals that a major fraction of high-z blazars are LSP objects. Their HE hump peaks at MeV energies and they 
		have higher Compton dominance than low-z blazars. The logarithmic histogram distributions of rest frame synchrotron peak frequency ($\nu_{syn}$), rest 
		frame IC peak frequency ($\nu_{IC}$), and Compton dominance (CD) roughly follow a Gaussian profile with $\mu \pm \sigma$ values of 
		log $ \nu_{syn} = 13.42 \pm 1.04 (15.32 \pm 1.46)$, log $ \nu_{IC}  = 22.00 \pm 1.07  (23.53 \pm 1.68)$  and log $ CD  = 0.48 \pm 0.50 \pm  (-0.4 \pm 0.40)$
		for high (low) redshift blazars. Also, the Pearson correlation study indicates that log $\nu_{syn} $ and log $\nu_{IC}$ are in anti-correlation with 
		with log (1 + z). log CD shows a strong positive correlation with log (1 + z). This is in accordance with the \emph{blazar sequence}. High-z 
		blazars, mostly being FSRQs, are intrinsically more luminous and therefore have high Compton dominance.
	
	\item   The SMBH mass distribution study of the blazar central black holes suggests that the average mass of SMBHs for high and low-z blazars is the same with 	 
		$\langle \text{log }M_{BH} \rangle = 8.6 $. Low-z blazars show a weak anti-correlation with log (1 + z) whereas high-z blazars have a mild positive 
		correlation with log (1 + z). Also, the average accretion disk luminosity and Eddington ratio of high (low) redshift blazars are 
		$\langle L_{Disk} \rangle = 5.62 \times 10^{45} \text{erg s}^{-1}(2.4 \times 10^{44} \text{erg s}^{-1})$,  $\langle \lambda_{Edd} \rangle = 0.12 (0.03)$ 
		with both the parameters showing a positive correlation with \mbox{log (1 + z).} 

	\item	We also estimated the masses of black hole seed (M$_{BH,seed}$) required for the formation of currently observed SMBHs in blazars using 
		simple exponential growth theory for the formation of SMBHs from the black hole seed via the accretion process. With $\eta=0.1 (0.3)$ and currently 
		observed constant $\lambda_{Edd}$, it is found that 26\% (69\%) of the total blazar population requires more massive seed progenitors than 
		what the direct collapse models or stellar dynamical processes predict. This implies that low accretion efficiencies and  high 	$\lambda_{Edd}$ 
		(sometimes super Eddington) episodes are required to explain the growth of observed SMBHs. It is also noticed that most of the BL Lacs (75\% and 92\% 
		for $\eta$ = 0.1 and 0.3, respectively) are required to have higher $\lambda_{Edd}$ than currently observed $\lambda_{Edd}$ to explain the growth of 
		their SMBHs. Therefore, BL Lacs should have had radiatively efficient accretion disk (FSRQ phase) during their lifetime before evolving into BL Lacs 
		having radiatively inefficient accretion disk.

	\item   According to the single zone leptonic emission model, emission zones of high (low) redshift blazars are required to have magnetic fields 
		$\langle \text {B } \rangle = 4.46G (1.09)$ G and electron energy distributions with $\gamma_p$ values $ \langle \gamma_p \rangle =7.94\times 10^2$ 
		(6.17 $\times 10^2$) to explain their broadband SEDs. They are broadly in agreement with the widely used blazar emission models. High-z blazars 
		(mostly FSRQs) contain low-energy photon-rich environments and relativistic electrons quickly lose their energies through IC scattering. Therefore, 
		their distribution has low $\gamma_p$. On the other hand, low-z blazars (mostly BL Lacs)  have a low-energy photon-starved environment surrounding them 
		and relativistic electrons can attain higher $\gamma_p$. This also contributes to the observed distribution of synchrotron and IC peak frequencies in 
		blazar SEDs. 
\end{itemize} 

Therefore, high-z blazars are a unique class of objects in the distant Universe. Their further detection and detailed broadband emission study will shed more light on
radiative and particle acceleration mechanisms under extreme conditions, cosmic evolution of SMBHs and relativistic jets, and intensity of EBL.  
Deep X-ray and $\gamma$-ray surveys are required to have an unbiased large population of blazars for further study of high-z blazars. In the coming decade,
the extended ROentgen Survey with an Imaging Telescope Array (eROSITA~\cite{Merloni2012}), the Large High Altitude Air Shower Observatory (LHAASO~\cite{Bai2019}),
the Southern Wide-field Gamma-ray Observatory (SWGO~\cite{Albert2019}), and the Cherenkov Telescope Array (CTA~\cite{CTA2021}) are expected to detect large 
number of high redshift blazars. 

\vspace{6pt}
%%%%%%%%%%%%%%%%%%%%%%%%%%%%%%%%%%%%%%%%%%
\authorcontributions{A.T.---Conceptualization,data curation, software and original draft preparation; K.K.S.---Conceptualization, review and editing; K.K.Y.---review and editing. All authors have read and agreed to the published version of the manuscript.}
\funding{This research received no external funding.}
\dataavailability{The data set used in this work is derived from the literature and can be shared on reasonable request.}
\acknowledgments{Authors thank all three anonymous reviewers for their important comments and suggestions. We also thank Hubing Xiao A. for sharing the blazar 
emission region data used in this study.}
\conflictsofinterest{The authors declare no conflicts of interest.} 
%------------------------------------References-------------------------------------------------------------
\begin{adjustwidth}{-\extralength}{0cm}

\reftitle{References}

\PublishersNote{}
\end{adjustwidth}
\end{document}